\documentclass[12pt,preprint]{aastex}
\usepackage{graphicx}
\usepackage{epsf}

\def\linebreak{\hfil\break}

\def\
{\hfil\linebreak}

\def\orch{{\it Orchestra}}
\def\deg{\ifmmode {^\circ}\else {$^\circ$}\fi}
\def\degree{\ifmmode {^\circ}\else {$^\circ$}\fi}
\def\mum{\ifmmode {\rm \mu {\rm m}}\else $\rm \mu {\rm m}$\fi}
\def\arcsec{\ifmmode ^{\prime \prime}\else $^{\prime \prime}$\fi}

\def\inch{\ifmmode ^{\prime \prime}\else $^{\prime \prime}$\fi}
\def\arcmin{\ifmmode ^{\prime}\else $^{\prime}$\fi}
\def\rfast{\ifmmode R_{fast}\else $R_{fast}$\fi}
\def\rgap{\ifmmode R_{gap}\else $R_{gap}$\fi}
\def\rhill{\ifmmode R_H\else $R_H$\fi}
\def\md{\ifmmode M_d\else $M_d$\fi}
\def\mp{\ifmmode M_P\else $M_P$\fi}
\def\mc{\ifmmode M_C\else $M_C$\fi}
\def\mh{\ifmmode M_H\else $M_H$\fi}
\def\mk{\ifmmode M_K\else $M_K$\fi}
\def\ms{\ifmmode M_S\else $M_S$\fi}
\def\mn{\ifmmode M_N\else $M_N$\fi}
\def\rp{\ifmmode R_P\else $R_P$\fi}
\def\rc{\ifmmode R_C\else $R_C$\fi}
\def\tpc{\ifmmode T_{PC}\else $T_{PC}$\fi}
\def\ppc{\ifmmode P_{PC}\else $P_{PC}$\fi}
\def\an{\ifmmode a_{Nix}\else $a_{Nix}$\fi}
\def\ak{\ifmmode a_{Kerberos}\else $a_{Kerberos}$\fi}
\def\as{\ifmmode a_{Styx}\else $a_{Styx}$\fi}
\def\ah{\ifmmode a_{Hydra}\else $a_{Hydra}$\fi}
\def\apc{\ifmmode a_{PC}\else $a_{PC}$\fi}
\def\aapc{\ifmmode {a_{PC} \over 5~\rp} \else $a_{PC}/5~\rp$\fi}
\def\epc{\ifmmode e_{PC}\else $e_{PC}$\fi}
\def\mpc{\ifmmode M_{PC}\else $M_{PC}$\fi}
\def\qpc{\ifmmode q_{PC}\else $q_{PC}$\fi}
\def\ompc{\ifmmode \Omega_{PC}\else $\Omega_{PC}$\fi}
\def\gyr{\ifmmode {\rm g~yr^{-1}}\else ${\rm g~yr^{-1}}$\fi}
\def\gcms{\ifmmode {\rm g~cm^{-2}}\else ${\rm g~cm^{-2}}$\fi}
\def\gcmc{\ifmmode {\rm g~cm^{-3}}\else ${\rm g~cm^{-3}}$\fi}
\def\R0{\ifmmode R_0\else $R_0$\fi}

\def\2470{[24]--[70]}

\def\cmsec{\ifmmode {\rm cm~s^{-1}}\else ${\rm cm~s^{-1}}$\fi}

\def\xm{$x_m$}
\def\pc{Pluto--Charon}

\newbox\grsign \setbox\grsign=\hbox{$>$} \newdimen\grdimen \grdimen=\ht\grsign
\newbox\simlessbox \newbox\simgreatbox
\setbox\simgreatbox=\hbox{\raise.5ex\hbox{$>$}\llap
     {\lower.5ex\hbox{$\sim$}}}\ht1=\grdimen\dp1=0pt
\setbox\simlessbox=\hbox{\raise.5ex\hbox{$<$}\llap
     {\lower.5ex\hbox{$\sim$}}}\ht2=\grdimen\dp2=0pt

\begin{document}

\title{The Formation of Pluto's Low Mass Satellites}
\vskip 7ex
\author{Scott J. Kenyon}
\affil{Smithsonian Astrophysical Observatory,
60 Garden Street, Cambridge, MA 02138} 
\email{e-mail: skenyon@cfa.harvard.edu}

\author{Benjamin C. Bromley}
\affil{Department of Physics, University of Utah, 
201 JFB, Salt Lake City, UT 84112} 
\email{e-mail: bromley@physics.utah.edu}
%
%

\begin{abstract}

Motivated by the {\it New Horizons} mission, we consider how Pluto's small 
satellites -- currently Styx, Nix, Kerberos, and Hydra -- grow in debris from the 
giant impact that forms the \pc\ binary.  After the impact, Pluto and Charon 
accrete some of the debris and eject the rest from the binary orbit. During 
the ejection, high velocity collisions among debris particles produce a 
collisional cascade, leading to the ejection of some debris from the system 
and enabling the remaining debris particles to find stable orbits around the 
binary.  Our numerical simulations of coagulation and migration show that 
collisional evolution within a ring or a disk of debris leads to a few small 
satellites orbiting \pc.  These simulations are the first to demonstrate 
migration-induced mergers within a particle disk. The final satellite masses 
correlate with the initial disk mass.  More massive disks tend to produce 
fewer satellites. For the current properties of the satellites, our results 
strongly favor initial debris masses of $3-10 \times 10^{19}$~g and current 
satellite albedos $A \approx$ 0.4--1.  We also predict an ensemble of smaller 
satellites, $R \lesssim$ 1--3~km, and very small particles, $R \approx$ 
1--100~cm and optical depth $\tau \lesssim 10^{-10}$.  These objects should 
have semimajor axes outside the current orbit of Hydra.

\end{abstract}


\section{INTRODUCTION}
\label{sec: intro}

The Pluto--Charon binary is an icy jewel of the solar system.  Discovered 
in 1978 \citep[][and references therein]{christy1978,nollSSBN}, the central 
binary has a mass ratio, $\qpc \equiv \mc\ / (\mp\ + \mc)$ = 0.10 
\citep[where \mp\ $\approx 1.3 \times 10^{25}$~g is the mass of Pluto and 
\mc\ $\approx 1.5 \times 10^{24}$~g is the mass of Charon;][]{buie2006}, 
an orbital semimajor axis \apc\ $\approx$ 17~\rp\ \citep[Pluto radii; 
1~\rp\ $\approx$ 1160~km;][]{young1994,young2007}, 
and an orbital period \tpc\ $\approx$ 6.4~d \citep[e.g.,][]{sicardy2006,buie2006}.
Four satellites -- Styx, Nix, Kerberos, and Hydra -- orbit at semimajor axes,
\as\ $\approx$ 37 \rp, \an\ $\approx$ 43~\rp, \ak\ $\approx$ 50~\rp, and 
\ah\ $\approx$ 57~\rp\ \citep{weaver2006,showalter2011,showalter2012,showalter2013}.
Curiously, the orbital periods of the satellites are roughly 3 (Styx), 4 (Nix), 
5 (Kerberos), and 6 (Hydra) times the orbital period of Charon. 
Although the satellite masses are uncertain by at least an order of magnitude
\citep[e.g.,][]{buie2006,tholen2008},
they are much less than the mass of the \pc\ binary. For an albedo $A$ and 
a mass density $\rho$ = 1 g cm$^{-3}$, the masses are 
\ms\ $\approx$ $9 \times 10^{16} A^{-3/2}$~g,
\mn\ $\approx$ $8.4 \times 10^{18} A^{-3/2}$~g,
\mk\ $\approx$ $2.6 \times 10^{17} A^{-3/2}$~g, and
\mh\ $\approx$ $1.4 \times 10^{19} A^{-3/2}$~g 
\citep[e.g.,][]{buie2006,showalter2011,youdin2012,showalter2012}.
For $A \approx$ 0.04--1 \citep[e.g.,][]{marcialis1992,roush1996,stansberry2008,brucker2009},
the combined mass of the four small satellites is $M_s \approx$ 
3--300~$\times~10^{19}$~g $\approx 2.3 - 230 \times 10^{-6}$~\mp. 

The architecture of \pc\ and the four smaller satellites provides fascinating tests
for the dynamical stability of multiple systems 
\citep[e.g.,][]{lee2006,tholen2008,suli2009,winter2010,peale2011,youdin2012}.  Dynamical 
fits of orbits to observations yield constraints on the masses, radii, and albedos of 
the satellites. Analyses using the restricted three-body problem or $n$-body simulations 
identify stable orbits surrounding \pc\ and place additional constraints on the masses 
of the satellites. Taken together, current theoretical results for Nix, Kerberos, and Hydra
suggest masses (albedos) close to the lower (upper) limits derived from dynamical fits 
to observations, with $M_s \lesssim 2.5 \times 10^{20}$~g and $A \gtrsim$ 0.2 
\citep{youdin2012}.

The system also challenges planet formation theories. Currently, most ideas center 
on a giant impact scenario, where the \pc\ binary forms during a collision between 
two icy objects with masses $M_1 \approx M_P$ and $M_2 \approx q M_P / (1 - q ) $, 
where $q \approx$ 0.25--0.33 \citep{mckinnon1989,canup2005,canup2011}.  
Immediately after the impact, the binary has an initial separation of 5--10 \rp.
Over 1--10 Myr, tidal evolution synchronizes the rotation of \pc\ and then 
circularizes and expands the binary to its present separation 
\citep{farinella1979,dobro1997}.

In some scenarios, the small satellites also form during the giant impact
\citep[e.g.,][]{stern2006,canup2011}.
Prior to the discovery of Kerberos/Styx, \citet{ward2006} developed an elegant model 
where the satellites move outward during the expansion of the \pc\ orbit. 
If the \pc\ orbit is eccentric throughout the expansion, resonant interactions 
with Charon can maintain the 1:4:6 period ratio of Charon, Nix, and Hydra.
\citet{ward2006} comment that a lower mass satellite might orbit with a 5:1 
period ratio. However, they also note that most of the impact debris probably 
developed high eccentricity orbits. High velocity collisions of debris particles 
complicate the ability of Charon to maintain a 3:4:5:6 period ratio for four satellites.

\citet{lith2008b} consider another path for satellite formation. Their analytic 
and numerical calculations suggest that maintaining the 4:1 (Nix) and 6:1 (Hydra) 
period ratios in an expanding binary requires mutually exclusive eccentricity 
evolution for the \pc\ binary \citep[see also][]{peale2011}.  Tidal interactions 
between Nix and Charon might establish the 4:1 period ratio, while interactions 
between Nix and Hydra establish the 4:6 period ratio. However, they consider this 
possibility unlikely. Instead, they propose that the \pc\ binary captured many 
small planetesimals into a disk. Collisional evolution then led to the formation 
of Styx, Nix, Kerberos, and Hydra; orbital migration later drove the orbits into their 
current configuration.

Here, we consider whether the \pc\ small satellites grow approximately {\it in situ} 
from material ejected during the impact.  We begin by demonstrating that a giant 
impact capable of forming Pluto--Charon is a natural outcome of scenarios for planet 
formation at 15--40 AU. After the impact, collisions among smaller objects orbiting 
Pluto-Charon produce a disk or a ring of debris.  For typical debris masses 
$M_d \approx 10^{19}$--$10^{21}$~g, collisions produce several satellites with 
masses comparable to those of Styx--Hydra at semimajor axes reasonably close 
to the current orbits of the satellites. Once small satellites form, they may 
migrate through the leftover debris. These results support the idea that satellites 
with masses comparable to those of Styx, Nix, Kerberos, and Hydra grow out of 
material in a circumbinary disk.

We outline the steps involved in this picture in \S2.  In \S3, we derive quantitative
constraints on (i) the probability of giant impacts (\S3.1),
(ii) the likelihood that Pluto was surrounded by a cloud of small objects prior to impact (\S3.2), 
(iii) the spreading time for the circumbinary ring (\S3.3),
(iv) time scales and possible outcomes for satellite growth within the ring (\S3.4), and
(v) plausible migration scenarios (\S3.5).
In \S4, we consider the implications of our results for the {\it New Horizons} mission \citep{stern2008}.
We conclude with a brief summary in \S5.  To isolate basic conclusions from descriptions 
of the numerical simulations, we use short summary paragraphs throughout the text. 

\section{PHYSICAL MODEL}
\label{sec: model}

To construct a predictive theory for the formation and evolution of satellite formation
around the \pc\ binary, we focus on the giant impact scenario. In this picture, two massive 
protoplanets within the protoplanetary\footnote{Throughout this paper, we use `circumplanetary' 
(`circumbinary') to refer to material orbiting a single (binary) planet and 
`protoplanetary' to refer to material orbiting the Sun.} debris disk collide and 
produce a binary planet embedded in a ring of debris.  Aside from the basic idea 
of a giant impact, our discussion diverges from Earth--Moon models. Unlike the 
Earth-Moon event, 
(i) the impact of icy protoplanets does not form a disk of vapor and molten material, 
(ii) a single object with roughly the mass of Charon survives the impact, and 
(iii) debris from the impact lies well outside the Roche limit of Pluto 
\citep[][and references therein]{canup2005,canup2011}.

Our approach has some features in common with recent scenarios where coagulation in
material just outside Saturn's ring system leads to the formation of several of the
innermost moons \citep[e.g.,][]{charnoz2010,charnoz2011}. 
Unlike Saturn, the evolution of the \pc\ binary has a large impact on the formation
and dynamical evolution of the satellites \citep[e.g.,][]{ward2006,lith2008b}. 
The large velocity dispersion of debris from the \pc\ impact also complicates 
coagulation calculations compared to Saturn's rings, where the low velocity dispersion 
allows rapid growth once material escapes the Roche limit\footnote{For a discussion 
of satellite formation in a disk around Mars, see \citet{rosen2012}.} 
\citep{salmon2010,charnoz2011}.

Our calculations follow some aspects of the \citet{ward2010} method for deriving the 
formation and evolution of satellites within the circumplanetary disks of gas giant planets 
\citep[see also][]{lun1982,canup2002,mosque2003a,mosque2003b,mosque2010,sas2010,ogi2012}. 
Like \citet{ward2010}, 
we consider the evolution of a circumplanetary disk where small particles grow into 
satellites.  For satellites around a gas giant, \citet{ward2010} derive the evolution 
of a circumplanetary disk of gas and dust fed from the surrounding protoplanetary disk.  
They include the growth and migration of satellites within the gaseous disk. For \pc, 
there is no gaseous component of the disk. Thus, we consider the evolution of a pure 
particle disk \citep[see also][]{ruskol1961,ruskol1963,ruskol1972,estrada2006}. Instead 
of deriving the radial disk structure and the growth of satellites analytically, we use 
a hybrid coagulation/$n$-body code to follow the growth and migration of satellites as 
a function of radial distance from the central binary \citep[see also][]{ogi2012}. 

Although satellite formation is a continuous process within an evolving protoplanetary disk,
it is convenient to list the main steps along the path from the formation of proto-Pluto 
and proto-Charon to the growth of stable satellites orbiting the \pc\ binary.

\begin{enumerate}

\item Proto-Pluto Formation -- within the protoplanetary disk, objects with $R \approx$ 
500--1000~km grow from much smaller planetesimals at semimajor axes $a \approx$ 20--50 AU 
on time scales of 10--100~Myr \citep[e.g.,][]{kb2012}. Typical calculations produce several 
Pluto-mass objects per AU \citep[e.g.,][]{kb2008,kb2010}.

\item Long-term Accretion -- once 500--1000~km protoplanets form, they continue to attract
material from the protoplanetary disk. Protoplanets accrete some small objects 
\citep[e.g.,][]{kb2008,kb2010} and capture others \citep[e.g.,][]{ruskol1972,weiden2002,gold2002}.  
Once a few objects lie in bound or temporary orbits, they may interact with others passing close 
to the planet \citep[e.g.,][]{ruskol1963,durda2000,stern2009,pires2012}.  Although the 
amount of material acquired by these processes is uncertain, they probably add little to 
the overall mass of the planet.

\item Impact -- forming a binary planet similar to Pluto-Charon requires collisions with an 
impact parameter $b = {\rm sin}~\theta \approx$ 0.75--0.95, where $\theta$ is the impact angle 
\citep{canup2005,canup2011}. After the collision shears off some material from the more massive 
planet (Pluto), the secondary (Charon) has an orbit with $\apc \approx 5-10 ~ \rp$ and 
$\epc \approx$ 0.1--0.8 \citep{canup2011}. Additional icy debris with a total mass 
$\md \lesssim 10^{23}$~g lies at distances of 1--30~\rp\ from Pluto.

\item Circumbinary Ring Formation -- after the binary forms, Pluto and Charon attempt to 
accrete and to eject the debris. Stable orbits around the binary have minimum semimajor axes 
ranging from $a \approx$ 2.5~\apc\ for \epc\ = 0.1 to $a \approx$ 4.3 \apc\ for \epc\ = 0.8
\citep[see][]{holman1999}. To achieve these orbits, debris particles must gain angular momentum 
and lose orbital energy. If a low mass cloud of captured objects surrounds proto-Pluto prior 
to impact, interactions between this cloud and debris from the impact produces a disk or ring 
of debris surrounding \pc.  Without this cloud, interactions among the debris particles and 
between the debris and \pc\ leads to a somewhat smaller disk or ring. In either case, debris 
particles initially have large orbital $e$; high velocity collisions are destructive and remove 
material from the binary system. 

\item Satellite Growth -- as \pc\ clear debris from the vicinity of their orbit, destructive 
collisions and radiation pressure begin to deplete material from the circumbinary disk or ring. 
Gravitational scattering spreads material to larger semimajor axes. Eventually, collisional 
damping overcomes secular perturbations from the central binary; growth by mergers begins. 
Growth leads to an ensemble of small satellites with masses and orbits set by the initial 
properties of the circumbinary disk.

\item Satellite Migration -- once small satellites form, they try to clear their orbits by 
scattering smaller objects.  Gravitational scattering among smaller objects tries to fill 
the orbits of the larger objects. When the large objects dominate, they reduce the surface 
density along their orbits and increase the surface density at smaller and larger semimajor
axes.  These surface density enhancements are not smooth; thus, the large objects feel a torque 
from the small objects.  If the sum of all of the torques does not vanish, the large objects 
migrate radially inward or outward \citep[e.g.,][]{gold1982,ward1997,kirsh2009}. In the right
circumstances, migration rates can be as large as 3--10~\rp\ every 1000 yr \citep{bk2013}. 

\item Tidal Expansion -- over long time scales of 0.1--10~Myr, tidal interactions between 
Pluto and Charon modify the orbital period and the rotational periods \citep{dobro1997}. 
On similar time scales, interactions between the binary and the surrounding disk can also 
modify the orbital period \citep{lin1979b,lin1979a}. Tidal forces rapidly synchronize 
Charon's rotational period with the orbital period on a time scale, 
$t_{sync} \approx$ $10^2$--$10^4$~yr for \apc\ $\approx$ 5--10~\rp\ \citep[see also][]{peale1999}.  
From an initial semimajor axis $a_{PC} \approx$ 5--10~\rp, tidal forces expand the orbit 
to the current \apc\ = 17 \rp\ in 0.2--20~Myr \citep[see also][]{farinella1979,dobro1997}.  

\end{enumerate}

Each of these steps involves complex physical processes that influence satellite growth. 
Accurate modeling of the full sequence is well beyond analytic theory and current numerical 
simulation. Here, we adopt the results of previous calculations of planet formation in the
protoplanetary disk \citep[e.g.,][]{kenyon2002,kb2004a,kb2004c,kb2008,kb2010,kb2012}.  
This information allows us to estimate (i) the frequency of giant impacts (\S3.1), and
(ii) the amount of protoplanetary material orbiting Pluto prior to impact (\S3.2).

Understanding the next step in the evolution -- the formation and evolution of a 
circumbinary ring of icy material -- is challenging.  Using constraints from detailed 
SPH simulations \citep{canup2005,canup2011}, we consider several physical processes that 
shape the physical extent of a circumbinary ring and develop constraints on the likely 
properties of the ensemble of small particles in the ring (\S3.3).  We use these results
as input to detailed numerical simulations of satellite growth (\S3.4) and migration 
(\S3.5). 

With this discussion, we generate realistic physical models piecewise to trace a 
reasonable path from the formation of proto-Pluto and proto-Charon to an ensemble of 
satellites orbiting the \pc\ binary.  These models provide quantitative, testable 
predictions for the upcoming New Horizons flyby.

\section{CALCULATIONS}

To develop clear constraints on this picture, we consider satellite formation in
the context of a standard model for planet formation in the protoplanetary disk 
\citep[e.g.,][]{saf1969,green1984,weth1989,kok1996,weiden1997b,kl1998,naga2005}.
In this model, the surface density of solids follows a power-law
\begin{equation}
\Sigma (a) = \Sigma_0 x_m a^{-k} ~ ,
\label{eq: sigma}
\end{equation}
where $\Sigma_0$ is a normalization factor designed to yield a surface density
equivalent to the `minimum mass solar nebula' at 1 AU \citep{weiden1977b,hayashi1981}, 
$x_m$ is a scaling factor to allow a range of initial disk masses, and $k$ = 1--2.
For $a$ measured in AU, $\Sigma_0 = 30$~g~cm$^{-2}$ \citep[see][and references therein]{kb2010}.

When the protoplanetary disk first forms, the solids consist of small 0.1--10~\mum\ particles
within a much more massive gaseous disk \citep[e.g.,][]{youdin2012b}.  Eventually, the 
solids aggregate into km-sized or larger planetesimals \citep[e.g.,][]{chiang2010,you11a}.
Binary collisions among planetesimals produce larger and larger protoplanets throughout the 
disk. At 20--50 AU, this process yields a large ensemble of Pluto-mass objects within 50--100~Myr 
after the formation of the Sun \citep[e.g.,][]{stern1995,stcol1997a,kl1999a,kb2004c,kbod2008}.

\subsection{The Setup: Frequency of Giant Impacts}

When Pluto-mass protoplanets first form, they contain a small fraction of the mass in 
solid material.  As long as the mass in planetesimals is larger than the mass in 
protoplanets, dynamical friction between the two sets of objects dominates viscous 
stirring among protoplanets \citep{gold2004}. Thus, protoplanet orbital eccentricities 
remain small. Collisions among protoplanets are then very rare. When protoplanets 
contain more than half of the total mass, viscous stirring dominates dynamical 
friction. Protoplanet eccentricities increase; chaotic growth begins. During 
chaotic growth, protoplanets grow through giant impacts with other protoplanets 
\citep[e.g.,][]{chambers2001a,kominami2002,kb2006,chambers2013}. Giant impacts 
remain common until protoplanets clear their orbits of other protoplanets and 
leftover planetesimals. 

Outcomes of giant collisions among protoplanets depend on the impact parameter, 
$b$, and the impact velocity $v_{imp}$ \citep{asphaug2006,lein2010,canup2011}.
Usually, $v_{imp} \gtrsim v_{esc}$, where $v_{esc}$ is the mutual escape velocity
of a merged pair of protoplanets.  Head-on impacts with $b \approx$ 0 then produce 
mergers with little or no debris.  Grazing impacts with $b \approx$~1 yield two
icy planets on separate orbits and some debris.  When $b \approx$ 0.75--0.95, 
collisions often produce a binary planet -- similar to the Earth--Moon or 
Pluto--Charon system -- embedded in a disk or ring of debris. 

The frequency of collisions that produce a binary planet depends on the distribution
of possible impact parameters. During the late stages of evolution, the system of
protoplanets is flattened, with a vertical scale height larger than the Hill radius 
of a Pluto-mass planet \citep{weth1993,weiden1997b,chambers1998,kb2006,kb2010}.
Assuming all impact orientations in the plane of the disk are equally likely 
\citep[e.g.,][]{agnor1999,chambers2013}, the probability that a collision will 
produce a binary planet is roughly the probability of an impact with $b \approx$ 
0.75--0.95, $\sim$ 33\%.

\begin{figure}
\includegraphics[width=6.5in]{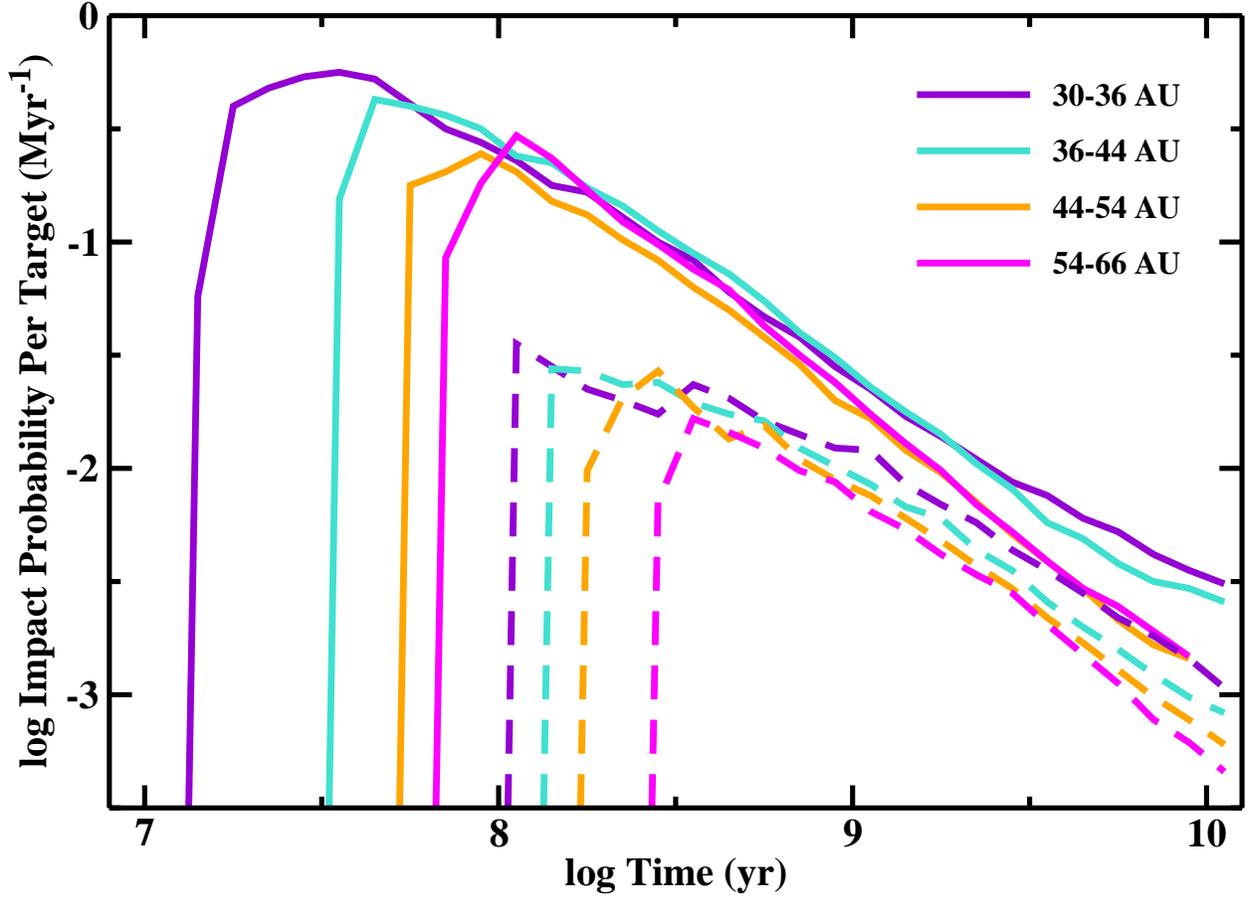}
\vskip 3ex
\caption{
Probability of giant impacts in protoplanetary disks with 
$\Sigma = \Sigma_0 x_m a^{-1}$. 
Solid curves: \xm\ = 0.33. Dashed curves: \xm\ = 0.10.
Giant impacts require collisions between a massive target, 
$m_t \gtrsim 10^{25}$ g, and a projectile with $m_p \gtrsim 0.3 m_t$.  
Colored curves illustrate the time evolution of the impact probability 
per target per Myr for disk annuli with boundaries listed in the legend.
Giant impacts (i) happen earlier in time closer to the star and 
(ii) occur earlier and more frequently in more massive disks.
\label{fig: imp1}
}
\end{figure}

To estimate the frequency of these impacts among icy objects during the early 
history of the solar system, we examine results from numerical calculations of 
planet formation at 15--150~AU around a solar-type star \citep{kb2008,kb2010,kb2012}. 
In these calculations, we follow the evolution of protoplanets and planetesimals as a 
function of semimajor axis $a$ and time $t$. Calculations begin with an initial 
ensemble of planetesimals with maximum radius \R0\ and roughly circular orbits 
in a disk with a surface density of solids,

Fig. \ref{fig: imp1} shows the frequency (number per target per Myr) of impacts 
with $b$ = 0--1 as a function of time for one set of calculations\footnote{The
figure illustrates results for ``strong'' planetesimals. Because our goal
is to derive initial estimates for the impact probability and the time 
between impacts, we defer a discussion of the variation of these parameters
as a function of the initial conditions and bulk properties of the planetesimals.} 
from \citet{kb2010}.  The curves plot the frequency of impacts between each target 
with mass $M_t \gtrsim 10^{25}$ g and a projectile with a mass 
$M_p \gtrsim q_p (M_t + M_p)$ where $q_p = 0.25$. As summarized in the caption, 
color indicates the semimajor axis of a disk annulus, ranging from 30--36 AU 
(violet) to 54--66 AU (magenta). Solid (dashed) lines show results for disks 
with $k$ = 1 and $x_m$ = 0.33 (0.10).

Throughout the evolution of the solar system, giant impacts are common 
\citep[see also][]{mckinnon1989,stern1991,canup2005,canup2011}.  Initially, 
all planetesimals are smaller than the target mass; the frequency of giant 
impacts is zero. As planetesimals grow into large protoplanets, the 
frequency of giant impacts increases. Because protoplanets grow faster 
in the inner disk than in the outer disk \citep{kb2004a,kb2004b}, giant 
impacts occur earlier in the inner disk (violet and turquoise curves) 
than in the outer disk \citep[orange and magenta; see also][]{liss1987,stern1995,kb2005}.  
In each disk annulus, the giant impact frequency per target rapidly rises 
to a clear peak when the surface density of targets and projectiles is 
largest.  As large objects collide and merge into larger protoplanets, 
the giant impact frequency declines. The rate of decline, 
$p_i \propto t^{-1}$, is the rate expected for collisional evolution 
in a protoplanetary disk \citep{kb2002a,dom2003,wya2008}.

Several factors contribute to the strong variation of $p_i$ with initial disk mass. 
In these calculations, the number of Pluto-mass objects $N_P$ grows with increasing 
disk mass, $N_P \propto x_m$ \citep{kb2008, kb2012}. For giant impacts requiring
two Pluto-mass objects, the number of possible pairs of impactors scales with the 
square of the disk mass.  In more massive disks, the larger mass of leftover 
planetesimals circularizes the orbits of the largest planets more effectively.  Thus, 
gravitational focusing factors also scale with disk mass. Combining these two
factors, the impact probability grows rapidly with disk mass, $p_i \propto x_m^k$,
with $k \approx$ 3--4.

To put these results in perspective, we define $t_N$ as the time to the next 
binary-producing giant impact on a single target.  If $p_i$ is the probability 
of a giant impact and if the distribution of impact orientations is random, 
the probability of an impact capable of producing a binary planet is 
$p_{bp} \approx 0.33 p_i$.  The time $t_N \approx p_{bp}^{-1}$. In a massive 
disk at 30--36~AU, $t_N \approx$ 5--10~Myr at an evolution time of 50--100 Myr. 
Binary planets then occur often, but they may be destroyed by another giant 
impact 5--10~Myr later. After an evolution time of 500~Myr to 1~Gyr (1--3~Gyr), 
$t_N \approx$ 150--300~Myr (500~Myr to 1~Gyr). Thus, binary icy planets form 
throughout the evolution of the planetary system.

\subsection{Before the Giant Impact: Long-Term Accretion}

After Pluto-mass objects form, they continue to interact with material from the
protoplanetary disk. Although protoplanets directly accrete some of this material,
they can also capture material onto temporary or bound orbits \citep[e.g.,][]{ruskol1972}.  
To estimate the amount of material Charon or Pluto capture in the 0.5--1~Gyr 
in between giant impacts, we consider a simple model. 
Objects from the protoplanetary disk enter the planet's Hill sphere at 
a rate $\dot{M}_H$. This material collides with other objects from the 
protoplanetary disk at a rate $\dot{M}_H \tau_{pd}$, where $\tau_{pd}$ is the optical depth 
of the protoplanetary disk. Collisions produce a single large remnant and some debris. The 
planet captures a fraction $f_c$ of this material into bound orbits.  Although other 
interactions are possible \citep[e.g.,][]{suet2011,pires2012,suet2013}, this approach 
yields a reasonable initial estimate for captured material 
\citep[e.g.,][]{ruskol1972,weiden2002,estrada2006}.

To derive the capture rate $\dot{M}_c$, we make several plausible assumptions. The optical depth of the 
protoplanetary disk is $\tau_{pd} \approx 3 \Sigma_{pd} / 4 \rho_{pd} R_{pd}$, where $\Sigma_{pd}$ is the
surface density and ($R_{pd}$, $\rho_{pd}$) are the radius and mass density of a typical object.
During the late stages of planet formation, $R_{pd} \approx$ 100~km \citep{kb2012}.  For
simplicity, we set $\rho_{pd}$ = 1 g~cm$^{-3}$. The protoplanetary disk has vertical scale 
height $H_z > R_H$ and surface density
$\Sigma_{pd} = d_{pd} \Sigma_0 x_m (a / {\rm 1~AU})^{-k}$ (see eq.~[\ref{eq: sigma}]),
where $\Sigma_0$ = 30~g~cm$^{-2}$, \xm\ = 1, $k$ = 1, and $d_{pd} < 1$ is a depletion factor 
which measures the fraction of disk material lost to collisional grinding 
\citep[e.g.,][]{kbod2008} and dynamical interactions \citep[e.g.,][]{morbiSSBN}.  
For epochs when giant impacts are common, $d_{pd} \approx$ 0.1.  Within the Hill sphere, 
collisions at $a \gtrsim \gamma R_H$ (with $\gamma \approx$ 0.3--0.4) produce no bound 
material \citep[e.g.,][]{toth1999,martin2011}. For collisions at smaller $a$, 
$f_c \approx$ 0.001 \citep[e.g.,][]{ruskol1972,weiden2002}.  

With these assumptions, disk objects enter the Hill sphere of the planet at a rate 
$\dot{M}_H \approx 3 \pi \Sigma_{pd} \Omega_{pd} (\gamma R_{H})^2$, where $\Omega_{pd}$ is 
the angular velocity of the planet around the Sun and the factor of three accounts 
for a small degree of gravitational focusing.  The capture rate is then
$\dot{M}_c \approx f_c \dot{M}_H \tau_{pd}$:
\begin{equation}
\dot{M_c} \approx 3 \times 10^{10} 
\left ( { d_{pd} \over 10^{-1} } \right )^2
\left ( { f_c \over 10^{-3} } \right ) 
\left ( { \gamma_c \over 0.333 } \right )^2 
\left ( { \Sigma_0 \over {\rm 30~g~cm^{-2}} } \right )^2 
\left ( { R \over \rp } \right )^2 
\left ( { R_{pd} \over {\rm 100~km}} \right )^{-1} 
\left ( { a \over {\rm 20~AU} } \right )^{-3/2} ~ {\rm g~yr^{-1}} .
\end{equation}
After $\sim$ 0.1--1~Gyr, modest reservoirs of material orbit Charon--Pluto mass planets.
Pluto captures 3--30 $\times 10^{18}$~g of orbiting debris. With a smaller radius and 
gravity, Charon attracts 1--10 $\times 10^{18}$~g. 

To estimate the collision frequency of captured material, we assume each fragment has radius
$R_f \sim$ 1~km, mass $M_f \sim 4 \times 10^{15}$~g, orbital semimajor axis 
$a_f \sim$~1000~\rp\ ($\sim 0.25 \rhill$), eccentricity $e_f \sim 0.1$, and inclination 
$i_f \sim \pi / 20$.  With $10^3 - 10^4$ fragments on randomly oriented orbits, the collision 
rate is $n_f \sigma_f v_f \approx$ 1--10 $(R_f / {\rm 1~km})^{-1}$~Gyr$^{-1}$, where $n_f$ 
is the number density, $\sigma_f$ is the cross-section, and $v_f$ is the relative velocity.  
In a cloud of 1~km objects, collisions are rare.  Smaller fragments collide more frequently.

To constrain the total specific angular momentum $L$ of the cloud, we set the specific angular 
momentum per fragment $L_f \approx (G \mp a_f (1 - e_f^2))^{1/2}$.  If each vector $\bf L_f$
is randomly oriented, the cloud has a typical $\bf L \approx$ 0 with a standard deviation of 
roughly $\sqrt{N} L_f$. Thus, there is a reasonable probability that a cloud of fragments has 
a net prograde or retrograde rotation around Pluto. 

Prior to impact, a low mass cloud of fragments probably surrounds Pluto. This mass depends on
the typical particle size and the depletion of the protoplanetary disk.  If collisions within 
Pluto's Hill sphere yield large fragments with $R_f \gtrsim$ 1~km, the fragments occupy a large 
volume, rarely collide, and have a small net prograde or retrograde angular momentum.  Despite
a potentially significant mass, this material makes little or no contribution to the formation 
of low mass satellites.

If collisions within Pluto's Hill sphere yield very small fragments, they collide more frequently 
and may produce a collisional cascade.  This process slowly reduces the circumplanetary mass.
With a small net prograde or retrograde $\bf L$, collisions may produce a disk of material with 
$a \ll a_f$ \citep[e.g.,][]{brah1976,brah1977}.  Although the properties of this disk are 
uncertain, debris from the \pc\ impact may interact with pre-existing disk material. 
Captured material may then contribute to the formation of low mass satellites.

For {\it New Horizons}, captured material at large $a$ is challenging to detect and an unlikely
hazard. Although 1~km fragments are detectable, they have small filling factor and are not
worth a detailed search. Within \pc's Hill sphere, the total collision probability between a
captured fragment and {\it New Horizons} is $\lesssim 10^{-13}$ throughout the flyby.

\subsection{After the Giant Impact: Dynamical Spreading of the Circumbinary Ring}

Shortly after the formation of the \pc\ binary, icy material with a total mass of
$\lesssim 10^{23}$~g lies at distances of 1--30~\rp\ \citep{canup2011}. As the evolution
proceeds, most of this material resides in a ring just outside the binary system.
To understand whether a circumbinary ring of debris might expand into a disk, we evaluate 
time scales for interactions among the ensemble of debris particles and between the binary 
and the debris. Particles of debris initially orbit with $a \approx$ 1--30~\rp\ \citep{canup2011}. 
The particles have radius $R$, mass density $\rho$ = 1 \gcmc, and total mass $\md$. SPH 
simulations of the \pc\ impact suggest $R \lesssim$~10~km and $\md \approx 10^{21} - 10^{23}$~g 
\citep{canup2011}. From more general simulations of impacts, the debris probably has a range of 
sizes from $\lesssim$ 0.1~km up to 10~km \citep[e.g.,][]{lein2012}.

Immediately after impact, collisions among debris particles produce a collisional cascade which
tends to circularize particle orbits, decrease particle radii, and remove mass from the system. 
Although the details of this process are uncertain, it probably leads to a ring of material with 
surface density $\Sigma$ at $a \approx 20~\rp$.  The \pc\ binary then has semimajor axis \apc\ = 
5--10~\rp, orbital period \tpc\ = 1--3~d, and eccentricity \epc\ $\approx$ 0.1--0.8. 

After the ring forms, two opposing mechanisms -- tides from the binary and collisional damping
among ring particles -- drive the evolution.  Tidal forces set the inner edge of the ring 
\citep[e.g.,][]{pri1991,holman1999} and impose a forced eccentricity, $e_f \approx 1.1 (\apc/a) \epc$, 
on ring particles \citep[e.g.,][]{hepp1974,murray1999,wyatt1999,mori2004,lee2006,tsuka2007}. 
Angular momentum transfer from the binary to the ring leads to a maximum eccentricity, 
$e_m \approx 2 e_f$ \citep[e.g.,][]{wyatt1999,mori2004} The time scale for secular perturbations 
from tides to raise $e$ is 
\begin{equation}
t_s \approx 300 \left ( { a \over 20 \rp } \right )^{7/2} \left ( 1 + {3 \over 2} \epc^2 \right )^{-1} \tpc 
\label{eq: t-tide}
\end{equation}
Aside from $a$, this time scale is independent of the properties of the ring. 
For \tpc\ = 1~d, $t_s \approx$ 1~yr. 

Physical collisions transfer angular momentum among ring particles \citep{gt78,horn1985}, 
which tends to circularize their orbits \citep[e.g.,][]{lbp1974,gt78}.  If $\Omega$ is 
the angular velocity of ring particles with number density $n$, cross-section $\sigma$, 
and relative velocity $v$, the time scale for physical collisions is 
$t_c \approx (dn/dt)^{-1} \approx (n \sigma v)^{-1} \approx \rho R / \Sigma \Omega$. Thus, 
\begin{equation}
t_c \approx 450 \left ( {R \over {\rm 1~km} } \right ) \left ( { a \over 20~\rp } \right )^{7/2} \left ( {10^{22}~{\rm g} \over \md } \right ) \tpc ~ .
\label{eq: t-coll}
\end{equation}
In an optically thick ring with $\md / R a^2 \gtrsim$ 1.5, the collision time is similar to or shorter 
than the secular time. 

When collisions are frequent, angular momentum transfer among ring particles slows the precession, 
prevents high velocity collisions, and lessens the impact of destructive collisions.  To conserve 
the angular momentum absorbed from the central binary, the ring must expand to larger 
$a$ \citep{lin1979b,lin1979a}.  Tides input angular momentum into the ring at a rate 
$\dot{L} \approx f \qpc^2 \apc^2 \ompc^2$, where $f$ is a constant of order unity 
\citep{lin1986}.  Setting the rate of angular momentum input equal to 
$\dot{L} = \dot{a} \Omega a / 2$, the expansion rate $\dot{a}$ for the ring is
\begin{equation}
\dot{a} \approx 5 ~ f ~ \left ( {a \over 20 \rp} \right )^{-1/2} \rp ~ {\rm yr^{-1}} ~ .
\label{eq: spread}
\end{equation}
The ring spreads from 20~\rp\ to 60~\rp\ in 5--10~yr, $\sim$ 20--40 collision times.

Although understanding the early evolution of the ring requires a detailed analytic 
\citep[e.g.,][]{raf2013} or numerical \citep[e.g.,][]{scholl2007,paarde2012} calculation, 
these estimates suggest a narrow ring of material plausibly expands into a disk at least
as fast as satellites grow from the debris (eq. [\ref{eq: t-coll}]; see \S3.4 below). For 
the very large collision velocities of particles with eccentricity $e_m$, destructive 
collisions can reduce typical particle sizes from $\sim$ 1~km to 0.1~km over 20--40 
collision times.  Significant mass removal on this time scale also seems likely. 

\subsection{Growth of Satellites in the Particle Disk}
\label{sec: sat-calc}

As the circumbinary ring of small particles expands, collisional damping reduces particle
random velocities. When random velocities are low enough, collisions produce merged objects
instead of a cloud of fragments. Dynamical friction lowers the random velocities of the
largest objects, promoting additional growth. Eventually satellites reach sizes of 1--10~km,
when they may begin to migrate through the disk.

To perform numerical calculations of planet and satellite formation, we use \orch, an 
ensemble of computer codes for the formation and evolution of planetary systems. Currently, 
\orch\ consists of a radial diffusion code which computes the time evolution of a gaseous 
or particulate disk \citep{bk2011a}, a multiannulus coagulation code which computes the 
time evolution of a swarm of planetesimals \citep{kb2004a,kb2008}, and an $n$-body code 
which computes the orbits of gravitationally interacting protoplanets 
\citep{bk2006,bk2011a,bk2013}.  The coagulation code uses statistical algorithms to evolve 
the mass and velocity distributions of low mass objects with time; the $n$-body algorithm 
follows the individual trajectories of massive objects.  Within the coagulation code, 
\orch\ includes algorithms for treating interactions between coagulation mass bins and 
the $n$-bodies \citep{bk2006}. To treat interactions between small particles and the 
$n$-bodies more rigorously, the $n$-body calculations include tracer particles. Massless 
tracer particles allow us to calculate the evolution of the orbits of small particles in 
response to the motions of massive protoplanets. Massive tracer particles enable calculations 
of the response of $n$-bodies to the changing gravitational potential of small particles
\citep{bk2011a, bk2011b, bk2013}.

We perform calculations on a cylindrical grid with inner radius $a_{in}$ and outer radius 
$a_{out}$. The model grid contains $K$ concentric annuli with widths $\delta a_i = 0.035 a_i$ 
centered at semimajor axes $a_i$. Calculations begin with a cumulative mass distribution 
$N_c(m_{ik}) \propto m_{ik}^{-q^{\prime}}$ of particles with mass density $\rho$ = 
1~g~cm$^{-3}$ and maximum initial mass $m_0$.  For comparison with investigators that 
quote differential size distributions with $N \propto m^{-q}$, $q^{\prime}$ = $q$ + 1.  
Here, we adopt an initial $q^{\prime}_0$ = 0.17; thus most of the initial mass is in the 
largest objects.

Planetesimals have horizontal and vertical velocities $h_{ik}(t)$ and $v_{ik}(t)$ relative 
to a circular orbit.  The horizontal velocity  is related to the orbital eccentricity, 
$e_{ik}^2(t)$ = 1.6 $(h_{ik}(t)/V_{K,i})^2$, where $V_{K,i}$ is the circular orbital 
velocity in annulus $i$.  The orbital inclination depends on the vertical velocity, 
$sin^2 i_{ik}(t)$ = $(2(v_{ik}(t)/V_{K,i})^2)$.

The mass and velocity distributions of the planetesimals evolve in time due to
inelastic collisions, drag forces, and gravitational encounters.  As summarized 
in \citet{kb2004a,kb2008}, we solve a coupled set of coagulation equations which
treats the outcomes of mutual collisions between all particles with mass $m_j$ 
in annuli $a_i$.  We adopt the particle-in-a-box algorithm, where the physical 
collision rate is $n \sigma v f_g$, $n$ is the number density of objects, $\sigma$ 
is the geometric cross-section, $v$ is the relative velocity, and $f_g$ is the 
gravitational focusing factor. Depending on physical conditions in the disk, we 
derive $f_g$ in the dispersion or the shear regime \citep{kb2012}.  For a specific 
mass bin, the solutions include terms for (i) loss of mass from mergers with other 
objects and (ii) gain of mass from collisional debris and mergers of smaller objects.

Collision outcomes depend on the ratio $Q_c/Q_D^*$, where $Q_D^*$ is the collision 
energy needed to eject half the mass of a pair of colliding planetesimals to infinity
and $Q_c$ is the center of mass collision energy.  From detailed $n$-body and smooth 
particle hydrodynamics calculations \citep{benz1999,lein2008,lein2009}, 
\begin{equation}
Q_D^* = Q_b R^{\beta_b} + Q_g \rho R^{\beta_g}
\label{eq: qdstar}
\end{equation}
where $Q_b R^{\beta_b}$ is the bulk component of the binding energy,
$Q_g \rho R^{\beta_g}$ is the gravity component of the binding energy, and $R$ 
is the radius of a planetesimal. We adopt
$Q_b$ = $2 \times 10^5$~erg~g$^{-1}$~cm$^{0.4}$, $\beta_b = -0.40$,
$Q_g$ = 0.22~erg~g$^{-2}$~cm$^{1.7}$, and $\beta_g$ = 1.30 
\citep{lein2008,lein2009}.

For two colliding planetesimals with masses $m_1$ and $m_2$, the mass of the merged 
planetesimal is
\begin{equation}
m = m_1 + m_2 - m_{ej} ~ ,
\label{eq:msum}
\end{equation}
where the mass of debris ejected in a collision is
\begin{equation}
m_{ej} = 0.5 ~ (m_1 + m_2) \left ( \frac{Q_c}{Q_D^*} \right)^{9/8} ~ .
\label{eq: mej}
\end{equation}
To place the debris in our grid of mass bins, we set the mass of the largest collision
fragment as $m_L = 0.2 m_{esc}$ and adopt a cumulative mass distribution
$N_c \propto m^{-q_d}$ with $q_d$ = 0.833, roughly the value expected for a system 
in collisional equilibrium \citep{dohn1969,will1994,tanaka1996,obrien2003,koba2010}. 
This approach allows us to derive ejected masses for catastrophic collisions
with $Q_c \sim Q_D^*$ and for cratering collisions with $Q_c \ll Q_D^*$
\citep[see also][]{weth1993,will1994,tanaka1996,stcol1997a,kl1999a,obrien2003,koba2010}. 

To compute the evolution of the velocity distribution, we include collisional damping 
from inelastic collisions and gravitational interactions.  For inelastic and elastic 
collisions, we follow the statistical, Fokker-Planck approaches of \citet{oht1992} and 
\citet{oht2002}, which treat pairwise interactions (e.g., dynamical friction and viscous 
stirring) between all objects in all annuli.  As in \citet{kb2001}, we add terms to 
treat the probability that objects in annulus $i_{k}$ interact with objects in annulus 
$i_{l}$ where $k, l$ = 1--$K$ \citep{kb2004a,kb2008}. We also compute long-range 
stirring from distant oligarchs \citep{weiden1989}. 

In the $n$-body code, we directly integrate the orbits of objects with masses larger
than a pre-set `promotion mass' $m_{pro}$. The calculations allow for mergers among
the $n$-bodies. Additional algorithms treat mass accretion from the coagulation grid 
and mutual gravitational stirring of $n$-bodies and mass batches in the coagulation 
grid. To treat interactions among proto-satellites, 
we set $m_{pro} = 10^{17} - 10^{19}$~g. 

Our calculations follow the time evolution of the mass and velocity distributions of 
objects with a range of radii, $r_{ij} = r_{min}$ to $r_{ij} = r_{max}$.  To track
the amount of material ejected by radiation pressure, we set $r_{min}$ = 20~\mum. 
The upper limit $r_{max}$ is always larger than the largest object in each annulus.  

For this initial exploration of satellite growth in a circumbinary ring, we derive
results for (i) pure coagulation calculations with fragmentation and no $n$-body 
component and (ii) hybrid calculations with the $n$-body component but no 
fragmentation. Aside from enabling a broader set of calculations per unit cpu time,
these two sets of calculations probably bracket the likely time scales for satellite
growth around the \pc\ binary. With no stirring from the central binary, the pure 
coagulation calculations set a firm lower limit on the growth time.  With no 
collisional damping or dynamical friction from small objects and no stirring from
the central binary, hybrid calculations without fragmentation likely establish an
upper limit on the growth time. Together, the two sets of calculations show that
the likely growth time is comparable to or longer than the time for the ring to 
spead (\S3.3).

We perform both sets of calculations on a grid of 32 concentric annuli extending 
from $a_{in}$ = 15~\rp\ to $a_{out}$ = 50~\rp.  This cylindrical grid provides a 
reasonably accurate representation of plausible orbits of debris particles from 
the giant impact around a circular binary system with an initial separation of 
5~\rp\ \citep{canup2011}.  The inner edge of this grid lies 
roughly at the innermost stable orbit around a newly-formed \pc\ binary. The outer 
edge allows the grid to cover a radial extent roughly two times larger than the 
current radial extent (39--57~\rp) of the satellite system. For a ring of fixed 
mass $M_d$ and aspect ratio $a_{out}/a_{in} \approx 3.33$, we expect satellites
to form at roughly the same relative positions with roughly the same mass on a
formation time which scales as $a_{in}^{2.5}$. Thus, choosing $a_{in}$ = 30~\rp\ and
$a_{out}$ = 100~\rp\ results in formation times that are roughly 5.5 times longer
than those described below.

\subsubsection{Pure Coagulation Calculations}

These calculations begin with a ring of material surrounding a single central object
with mass equivalent to the \pc\ binary. The total mass in the ring ranges between 
$M_d = 3 \times 10^{19}$~g and $M_d = 3 \times 10^{21}$~g, with an initial surface 
density distribution $\Sigma \propto a^{-1}$.  Within each annulus of the ring, the 
solid particles range in size from 20~\mum\ to 0.1~km. Radiation pressure probably 
ejects particles with $R \lesssim$ 20~\mum\ on short time scales 
\citep[e.g.,][]{burns1979,poppe2011}. Although particles with $R \gtrsim$ 0.1~km might
lie within the ring, adopting a maximum particle size of 0.1~km provides a rough upper 
limit on the amount of mass lost to a collisional cascade.  We consider calculations
starting with several 1~km objects in the next section.  To allow the system to find an 
equilibrium between the large orbital eccentricity expected of material ejected from the 
central binary and the small eccentricity expected from collisional damping, we adopt an 
initial eccentricity $e_0$ = 0.1 and an initial inclination $i_0/e_0$ = 0.5. 

All calculations follow a similar pattern. With the large initial $(e, i)$ of all particles, 
collisions among large particles are destructive and produce copious amounts of smaller 
particles. Among the smaller particles, collisional damping is effective. Through dynamical 
friction, smaller particles damp the orbits of the larger particles. Thus, the $(e, i)$ 
for all particles declines with time. Collisions become less destructive.  Eventually, 
collisional velocities become small enough to promote growth over destruction. During this
phase, roughly 25\% of the initial mass is ground down into particles smaller than 20~\mum.
Radiation pressure ejects this material from the binary \citep[e.g.,][]{burns1979,kw2011}.

After the system reaches a low velocity equilibrium, the largest particles begin to grow 
rapidly. During a very short phase of runaway growth, objects grow from roughly 0.1~km
to 3--30~km. As the largest objects grow, they gravitationally stir the smaller particles. 
Once the largest objects contain most of the mass, stirring dominates collisional damping.
The orbital $(e, i)$ of the smaller objects rapidly increases, reducing gravitational
focusing factors and halting runaway growth. At the end of runaway growth, each annulus 
contains a handful of proto-satellites with radii of 3--30~km. 

Following runaway growth, the evolution of the system stalls. The proto-satellites rapidly
accrete any small particles left in their annuli.  Because the small particles contain so
little mass, this growth of the large particles is very small.  In the particle-in-a-box
approximation adopted for these pure coagulation calculations, the probability of collisions 
among large objects is small. Thus, the sizes of the proto-satellites reach a rough steady 
state set approximately by the initial mass in each annulus of the grid.

\begin{figure}
\includegraphics[width=6.5in]{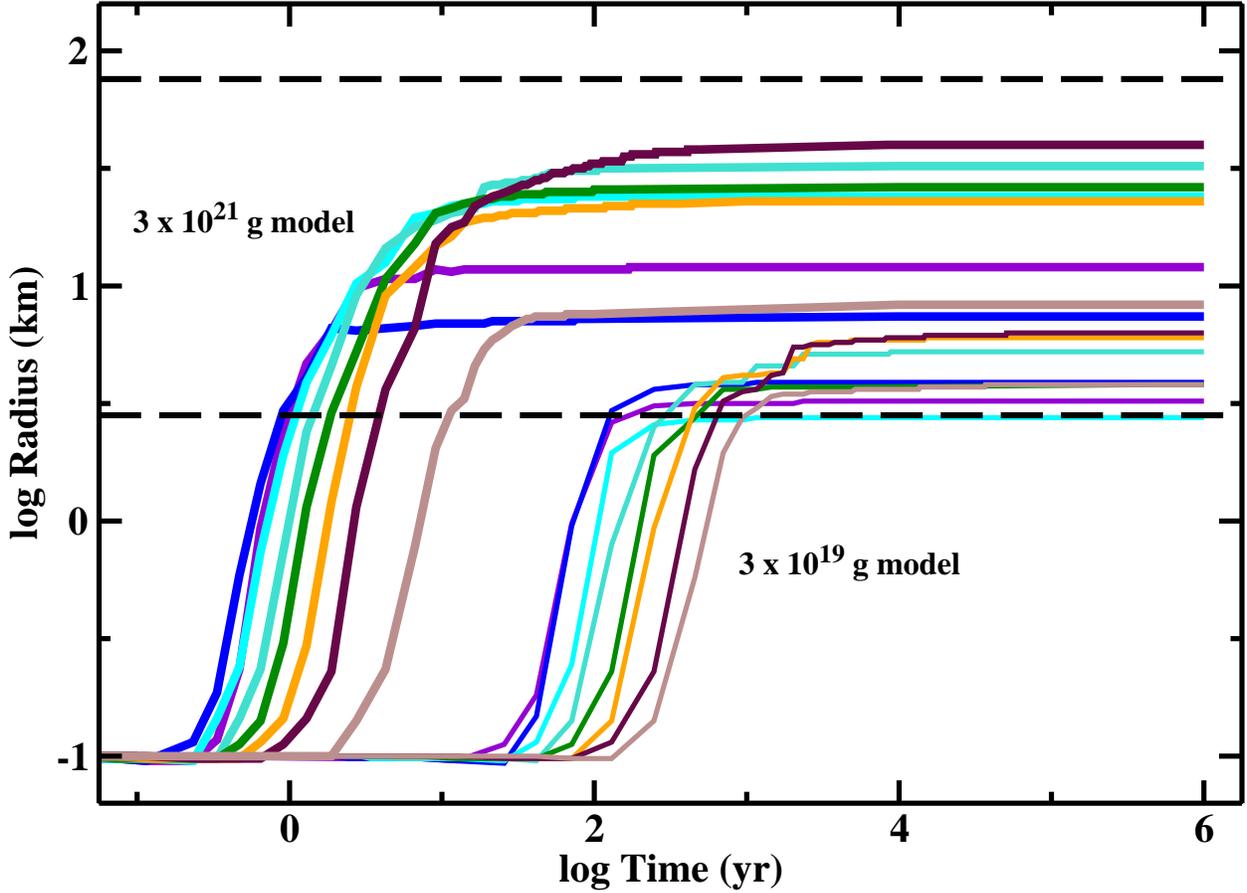}
\vskip 3ex
\caption{
Radius evolution for satellites in a disk surrounding the \pc\ binary. The dashed 
lines indicate the lower limit for the radius of Styx ($A$ = 1) and the upper limit 
for the radius of Hydra ($A$ = 0.04) assuming a mass density $\rho$ = 1 g~cm$^{-3}$.
The thin (thick) lines plot results for a model with an initial disk mass of
$M_d = 3 \times 10^{19}$~g ($M_d = 3 \times 10^{21}$~g) in solid objects with
initial radii of 0.1~km and smaller. Each line indicates the radius evolution for 
the largest object in annuli at 17.5~\rp\ (violet), 20~\rp\ (blue), 23~\rp\ (cyan), 
26.5~\rp\ (turquoise), 30.5~\rp\ (green), 35~\rp\ (orange), 40.5~\rp\ (maroon), 
and 46.5~\rp\ (brown). Satellites grow faster in more massive disks. The range of
model radii span the range inferred for satellites of \pc.
\label{fig: rgrow1}
}
\end{figure}

Figure \ref{fig: rgrow1} illustrates the growth of satellites for calculations with 
$M_d = 3 \times 10^{19}$~g 
(thin lines) and $M_d = 3 \times 10^{21}$~g (thick lines). As summarized in the caption,
each line represents the evolution in the radius of the largest object in annuli ranging
in distance from 15~\rp\ to 50~\rp\ from the center-of-mass. In these disks, satellite 
growth is very rapid. After an initial period of destructive collisions, it takes only 
1--100 yr for an ensemble of 0.1~km fragments to grow into 10~km satellites.  The growth 
time is inversely proportional to the initial disk mass, 
$t_g \approx 100 ~ (3 \times 10^{20}~{\rm g} / M_d ) ~ {\rm yr}$.
Once satellites reach a maximum radius of 5--50~km, growth stalls. Satellites remain at a 
roughly constant radius for thousands of years.

These calculations demonstrate the inevitable growth of 5--30~km satellites from an 
ensemble of much smaller objects. Independent of the initial orbital eccentricity, 
collisional damping reduces the internal velocity dispersion to levels that promote 
mergers during collisions. Although the collisional cascade removes roughly 25\% of 
the initial mass, small objects grow rapidly into satellites with radii of 5--30~km. 

Despite the efficiency of small satellite formation, coagulation calculations produce 
too many satellites. In these models, each of the 32 annuli in the calculation yields
0--3 satellites with radii of 5--30 km. Even with small number statistics, this result 
is much larger than the current number of known satellites. However, the coagulation
calculations do not allow large-scale dynamical interactions among satellites. For the
models shown in Fig. \ref{fig: rgrow1}, satellites in adjacent annuli have orbital 
separations of 5--15~\rp. Thus, satellites are close enough to perturb the orbits of 
their nearest neighbors, leading to chaotic interactions as in simulations of 
terrestrial planet formation \citep{chambers2001a,kok2002,kominami2004,kb2006}.

\subsubsection{Hybrid Calculations}

To investigate dynamical interactions among newly-formed satellites, we now consider a
suite of hybrid simulations with {\it Orchestra}. Objects in the coagulation code which
reach a preset mass, $m_{pro}$, are promoted into the $n$-body code. When a satellite with 
$(m_{ij}, e_{ij}, i_{ij})$ in an annulus with semimajor axis $a_i$ is promoted, we assign 
$(e_n, i_n)$ = $(e_{ij}, i_{ij})$, $a_n$ = $a_i + g \Delta a_i$, and a random orbital phase, 
where $g$ is a random number in the range $-0.5$ to $+0.5$. The $n$-body code converts these 
orbital elements into an initial $(x, y, z)$ position and an initial $(v_x, v_y, v_z)$
velocity vector.  The $n$-body code follows the trajectories of all promoted satellites. 
Algorithms within {\it Orchestra} allow $n$-bodies to interact with coagulation particles 
\citep{bk2006,bk2011a}.

The starting conditions for these calculations are identical to those for the pure 
coagulation models. We consider three cases for the initial disk mass, 
$m_d = 3 \times 10^{19}$~g (low mass disk), $m_d = 3 \times 10^{20}$~g (intermediate mass disk),
and $m_d = 3 \times 10^{21}$~g (massive disk).  For each case, we adopt a different promotion 
mass: $m_{pro} = 3 \times 10^{17}$~g (low mass), $m_{pro} = 10^{18}$~g 
(intermediate mass), and $m_{pro} = 3 \times 10^{18}$~g (high mass). These promotion masses 
are a compromise between the `optimal' promotion mass, $m_{pro} = 3-30 \times 10^{16}$~g, 
required to allow newly formed $n$-bodies to adjust their positions and velocities to existing 
$n$-bodies and the practical limits required to complete calculations in a reasonable amount 
of time.  For each combination of disk mass and promotion mass, we ran 20--25 simulations.  

To check the accuracy of our results for the intermediate and high mass disks, we ran an 
additional 12 simulations for each of two cases with promotion masses a factor of three 
larger and a factor of three smaller than the promotion masses listed above. When the 
promotion mass is a factor of three larger, the chaotic growth phase begins later and lasts 
longer. Often, chaotic growth in the outer disk is well-separated in time from chaotic 
growth in the inner disk. This unphysical behavior contrasts with calculations with smaller 
promotion masses, where the timing of chaotic growth changes smoothly from the inner disk
to the outer disk. As a result, these calculations have less scattering and radial mixing 
among proto-satellites. When the promotion mass is a factor of three smaller, the onset, 
character, and duration of chaotic growth changes very little.  In all cases, the final 
number of satellites is fairly independent of promotion mass. 

Because fragmentation has a small impact on the results of pure coagulation calculations
per unit cpu time, these hybrid calculations do not include fragmentation. Neglecting 
fragmentation increases the mass available for satellite growth by roughly 25\%. Without
fragmentation, collisions fail to produce copious amounts of 1~m to 100~m particles.
Collisional damping among these debris particles and dynamical friction between the
debris and larger survivors reduces the orbital $e$ and $i$, aiding runaway growth. Thus, 
these hybrid calculations artificially increase the evolution time required for satellites 
to reach $R \gtrsim$ 10~km.

\begin{figure}
\includegraphics[width=6.5in]{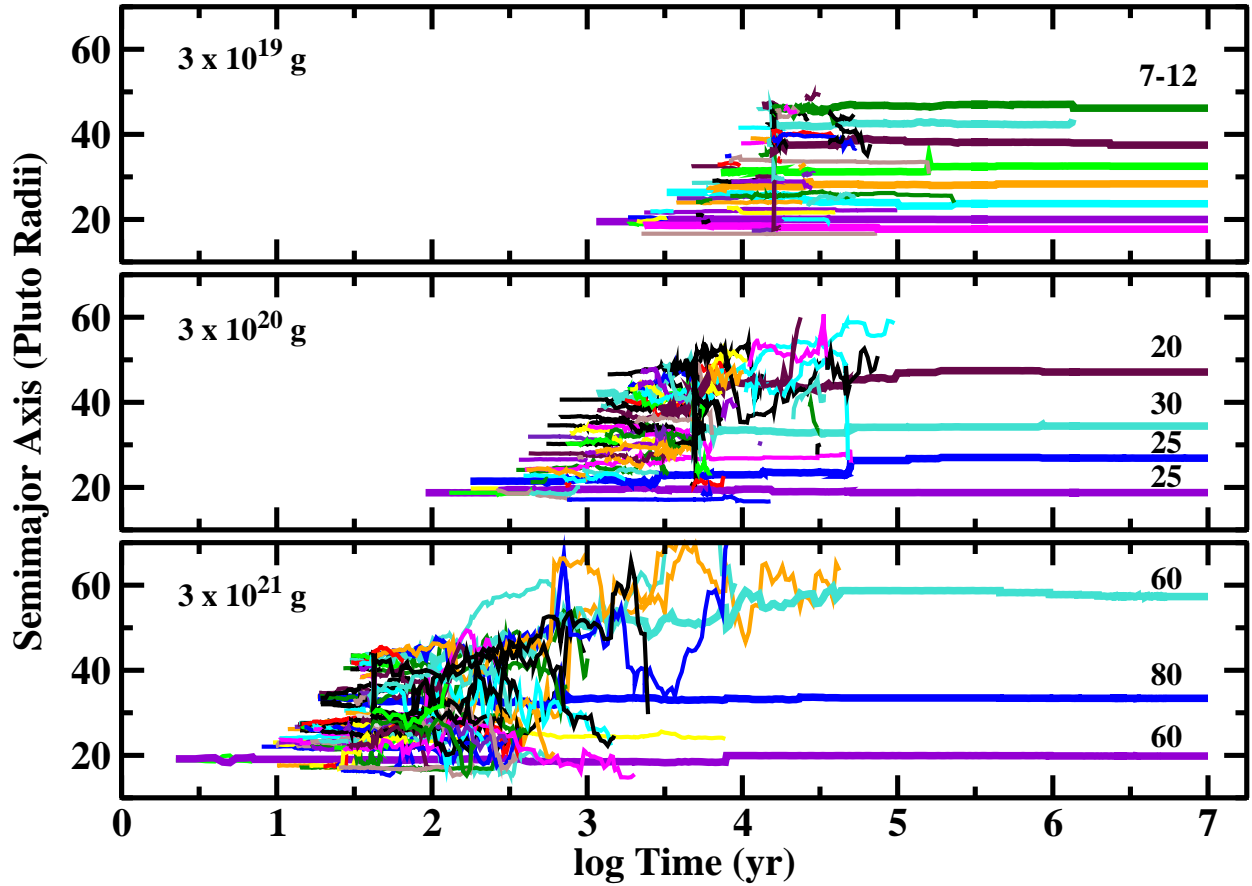}
\vskip 3ex
\caption{
Semimajor axis evolution for satellites in a disk surrounding \pc. The upper left 
corner of each panel lists the initial disk mass. Along the right edge of each
panel, numbers indicate the radii (in km) of each satellite. More massive disks
produce fewer, larger satellites than less massive disks.
\label{fig: rsemi1}
}
\end{figure}

Fig.~\ref{fig: rsemi1} shows the evolution of the semimajor axes for promoted satellites 
orbiting a single central object with the mass of the \pc\ binary. All objects begin with 
$r \lesssim$ 0.1~km ($m \lesssim 4 \times 10^{12}$~g).  The evolution time required for 
these objects to reach the promotion mass scales with the initial disk mass, 
$t_{pro} \approx 10^3 ~ (3 \times 10^{19}~{\rm g} / m_d)$~yr. This time scale is roughly a
factor of ten longer than derived for pure coagulation calculations with fragmentation.  In 
the inner disk, objects have shorter orbital periods and shorter collision times. Thus, the 
first promoted objects appear near the inner edge of the disk. As the calculation proceeds, 
satellites are promoted farther and farther out in the disk. Eventually, a few promoted 
objects appear in the outer disk.

Every calculation experiences chaotic growth \citep{gold2004,kb2006}. During chaotic growth, the
satellites scatter one another throughout the grid. Because objects grow fastest in the most
massive disks, chaotic growth begins earlier -- $\sim$ 100~yr -- in the high mass disk and much
later -- $\sim 10^4$~yr -- in the low mass disk. Chaotic growth is also `stronger' in more massive
disks. In more massive disks, massive large satellites are more numerous and generate larger 
radial excursions of smaller satellites than in lower mass disks.  

Throughout the evolution shown in Fig.~\ref{fig: rsemi1}, the $n$-bodies slowly accrete the
very small planetesimals remaining in the coagulation grid. The number of leftovers correlates
well with the number of $n$-bodies. Thus, the total mass in leftover planetesimals within the 
coagulation grid approaches zero as the number of $n$-bodies reaches a minimum. 

In all of these calculations, the final number of satellites correlates inversely with the 
initial disk mass. In massive disks with $M_d = 3 \times 10^{21}$~g, there are usually 2 or
3 massive satellites with $r \approx$ 50--80~km at the end of the calculation. As we reduce
the initial disk mass, the calculations produce more satellites with smaller masses. In
intermediate mass disks with $M_d = 3 \times 10^{20}$~km, chaotic growth leaves all of the
initial mass in 4--5 satellites with radii of 20--30~km. In low mass disks, there are very
few mergers during chaotic growth. After $\sim 10^7$~yr, there are typically 7--9 satellites
with radii of 7--12~km.  

To test how outcomes change with the initial size distribution, we consider a suite of 15 simulations
with two additional 10~km planetesimals placed randomly in the grid. These large planetesimals 
stir their surroundings and thus counteract collisional damping by nearby small planetesimals. 
With much larger masses than any other planetesimals in the grid, they have large collisional 
cross-sections and can grow very rapidly. 

\begin{figure}
\includegraphics[width=6.5in]{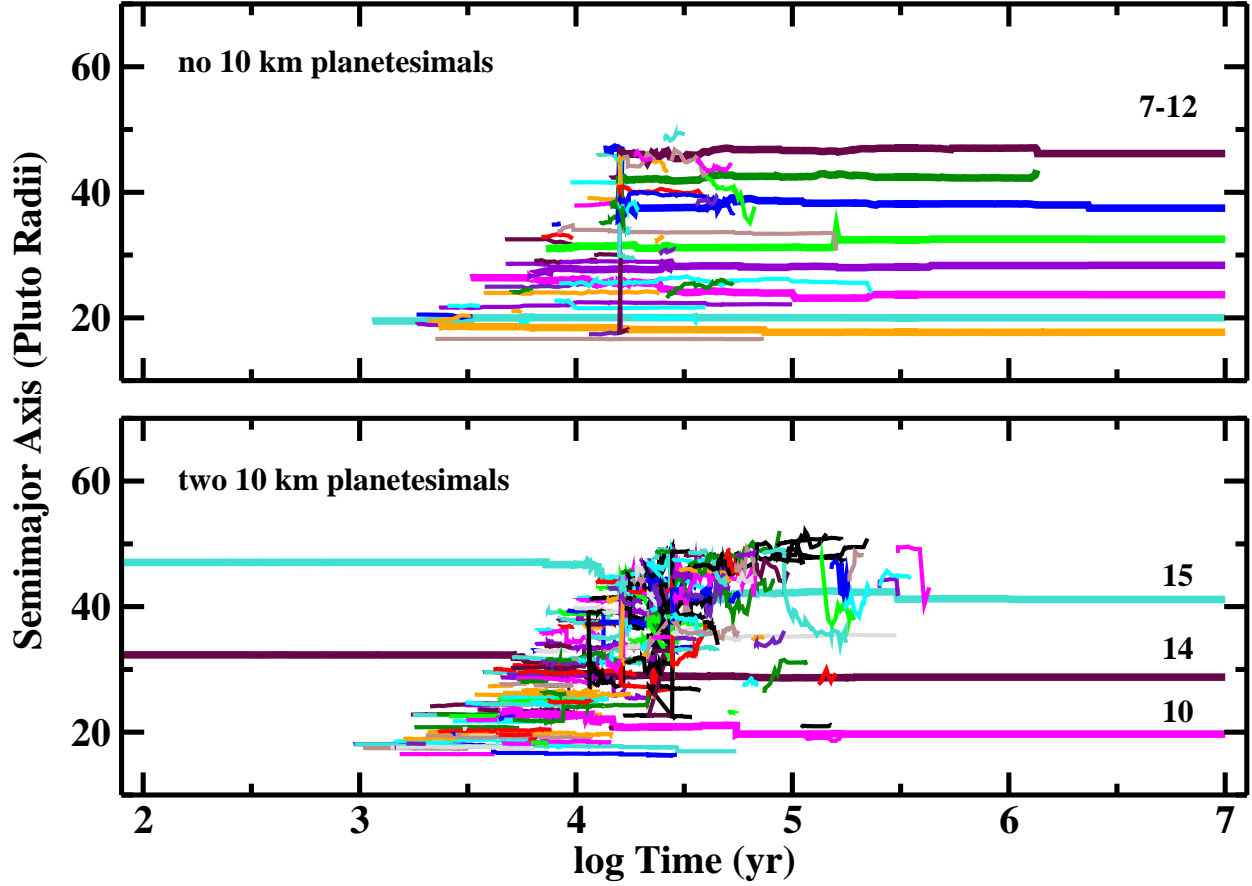}
\vskip 3ex
\caption{
As in Fig.~\ref{fig: rsemi1} for an initial disk mass 
$m_d \approx 3 \times 10^{19}$~g composed of 0.1~km and 
smaller planetesimals.
Upper panel: evolution without larger planetesimals;
lower panel: evolution with two 10~km planetesimals.
Calculations with a several initial large planetesimals 
yield fewer, larger satellites than calculations without
large planetesimals.
\label{fig: rsemi2}
}
\end{figure}

Fig.~\ref{fig: rsemi2} compares the time evolution of the semimajor axes for $n$-bodies in
calculations with (lower panel) and without (upper panel) two additional massive planetesimals.
In standard calculations with $M_d = 3 \times 10^{19}$~g, there is a short period of chaotic
growth at $\sim 10^4 - 10^5$~yr followed by a long quiescent period with an occasional merger.
When there are two large planetesimals at the onset of the calculation, these planetesimals 
slowly accumulate small planetesimals. Meanwhile, small planetesimals in the rest of the disk 
rapidly grow to sizes (4--5~km) that allow promotion into the $n$-body code. From Fig.~\ref{fig: rsemi2},
the time scale for these objects to reach the promotion mass is roughly $10^3 - 10^4$~yr 
independent of the two large satellites. However, stirring by the two large satellites promotes
an earlier oligarchic growth phase which enables more uniform growth of the largest planetesimals.
Thus, there are many more promoted objects and a more intense chaotic growth phase. In addition 
to the two initial massive planetesimals, one or two other planetesimals accrete all of the other 
promoted objects.  After $\sim 10^5$ yr, there are only a few small $n$-bodies left.  The 
3--4 massive satellites accrete these objects in a few hundred thousand years. 

This result is typical of all calculations in low or intermediate mass disks that begin with a few 
large (10~km) planetesimals placed randomly throughout the grid.  Stirring by the large objects slows
runaway growth and allows a larger group of oligarchs to grow rapidly.  Although these oligarchs 
reach the promotion mass on similar time scales, many more of them reach the promotion mass.  
Typically, calculations with initial large planetesimals produce 2--4 times as many $n$-bodies 
as calculations without large planetesimals. Stirring by the 10~km planetesimals and interactions 
among the many $n$-bodies leads to a very active phase of chaotic growth. During chaotic growth, 
the radial excursions of the $n$-bodies are 2--3 times larger in semimajor axis. Collisions among 
$n$-bodies are more frequent, leading to a system with fewer, but more massive, satellites. 
With a set of two initially large planetesimals, low (intermediate) mass disks produce 
3--4 (2--3) massive satellites instead of 7--9 (4--5).

\begin{figure}
\includegraphics[width=6.5in]{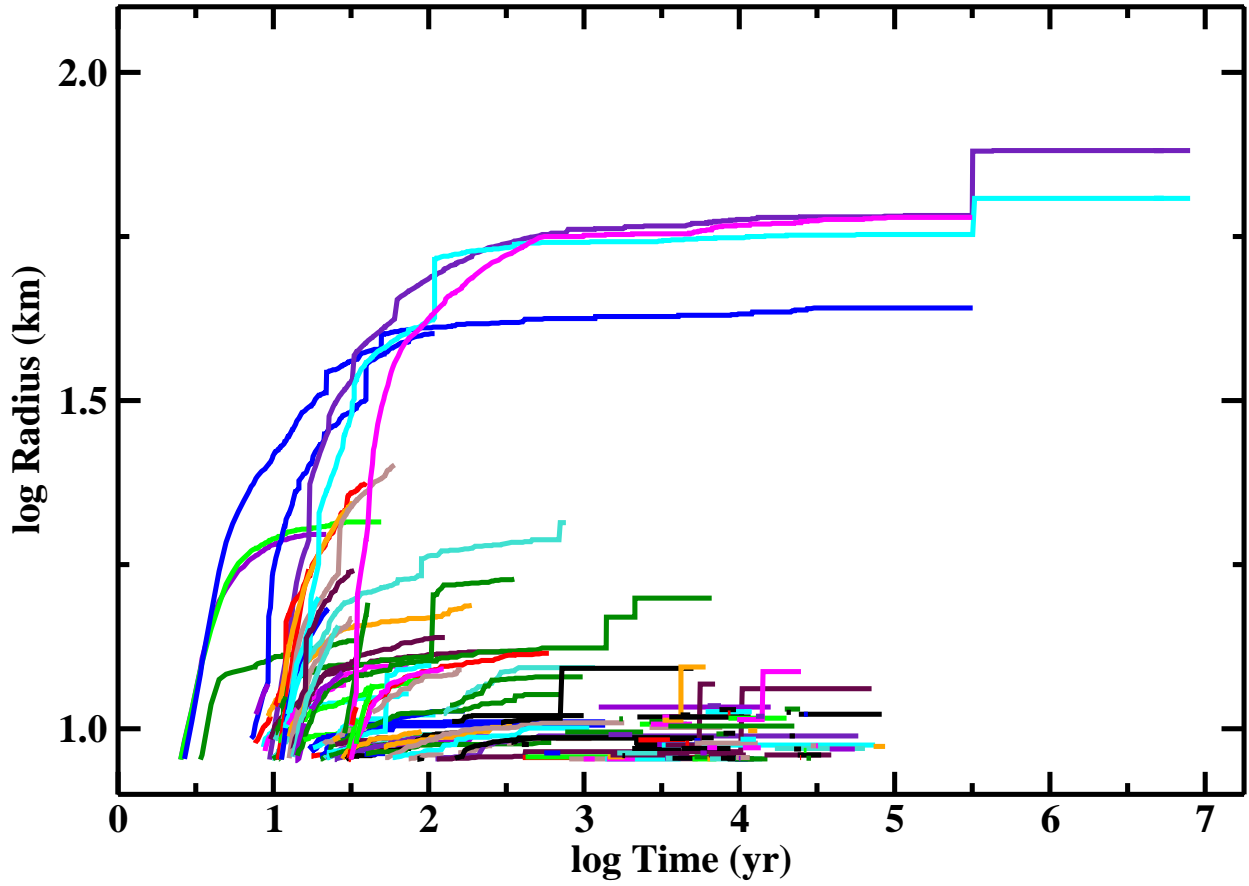}
\vskip 3ex
\caption{
Radius evolution for $n$-bodies. The calculation begins with
$m_d = 3 \times 10^{21}$~g and $m_{pro} = 3 \times 10^{18}$~g.
Once a few large objects form, they gradually accrete all of
the smaller $n$-bodies.
\label{fig: rgrow2}
}
\end{figure}

For all of these hybrid calculations, the radius evolution follows a standard pattern 
(Fig.~\ref{fig: rgrow2}).  As in Fig.~\ref{fig: rgrow1}, the largest objects in the coagulation grid 
grow from 0.1~km to 1--10~km. Once they are promoted into the $n$-body grid, they interact with 
objects in the coagulation grid {\it and} all of the promoted objects in the $n$-body grid.  Unlike 
the pure coagulation calculations, growth does not stall at 5--30~km. The extra interactions between
promoted objects and the larger volume sampled by scattered objects enables a few large objects to 
accumulate nearly all of the remaining mass in the grid. Thus, these objects reach radii that are 30\% to 
100\% larger (and 2--8 times more massive) than the largest objects in the pure coagulation calculations.

\subsubsection{Summary of Coagulation and Hybrid Calculations}

The suite of pure coagulation and hybrid calculations demonstrates that the \pc\ satellites 
grow rapidly from a disk of small planetesimals. In calculations with fragmentation, 
destructive collisions grind down 
roughly 25\% of the initial disk mass into 20~$\mu$m particles which are ejected from the 
binary system. Collisional damping among larger debris particles reduces orbital 
eccentricity of 0.1--10~m objects. Dynamical friction between these small planetesimals 
and the surviving large planetesimals damps the orbital eccentricities of the largest 
objects, greatly reducing the frequency of destructive collisions and enabling rapid 
growth of surviving planetesimals.  Thus, fragmentation removes mass and aids the eventual 
rapid growth of satellites.

In hybrid calculations without fragmentation, less collisional damping and dynamical friction
slow growth considerably. However, satellites still grow on time scales similar to the expansion
time for the central binary.  The number of satellites $N_s$, typical satellite radius $R_s$, 
and the time for satellites to reach their final mass $t_f$ scales with the initial mass of the 
planetesimal disk. For calculations without any large (10~km) fragments at $t = 0$ and 
$M_d \approx 3 - 300 \times 10^{19}$~g, we infer 
\begin{equation}
\left.
\begin{array}{l}
N_s \\
\\
R_s \\
\\
t_f \\
\end{array}
\right\}
=
\left\{
\begin{array}{ll}
2^{22.5 - {\rm log}~M_d} \\
\\
27.5 \left ( { 3 \times 10^{20}~{\rm g} \over M_d } \right )^{0.425} ~ {\rm km} & ~~~~~~~~~~~~~{\rm no ~ initial ~ large ~ fragments} \\
\\
10^4 \left ( { 3 \times 10^{20}~{\rm g} \over M_d } \right ) ~ {\rm yr}
\end{array}
\right.
\end{equation}

In low and intermediate mass disks, calculations with several 10~km fragments produce 
fewer satellites with larger masses. For $M_d \approx 3 - 30 \times 10^{19}$~g, 
$N_s \approx$ 3--5 and $R_s \approx$ 15--60~km instead of the $N_s \approx$ 4--10 
and $R_s \approx$ 7--30~km for calculations without large fragments. Because large 
fragments tend to sweep up any small planetesimals, calculations with fragments yield
a smaller spread in the number and masses of planetesimals than calculations without
fragments.

\subsubsection{Observations: Comparisons and Predictions}

\begin{figure}
\includegraphics[width=6.5in]{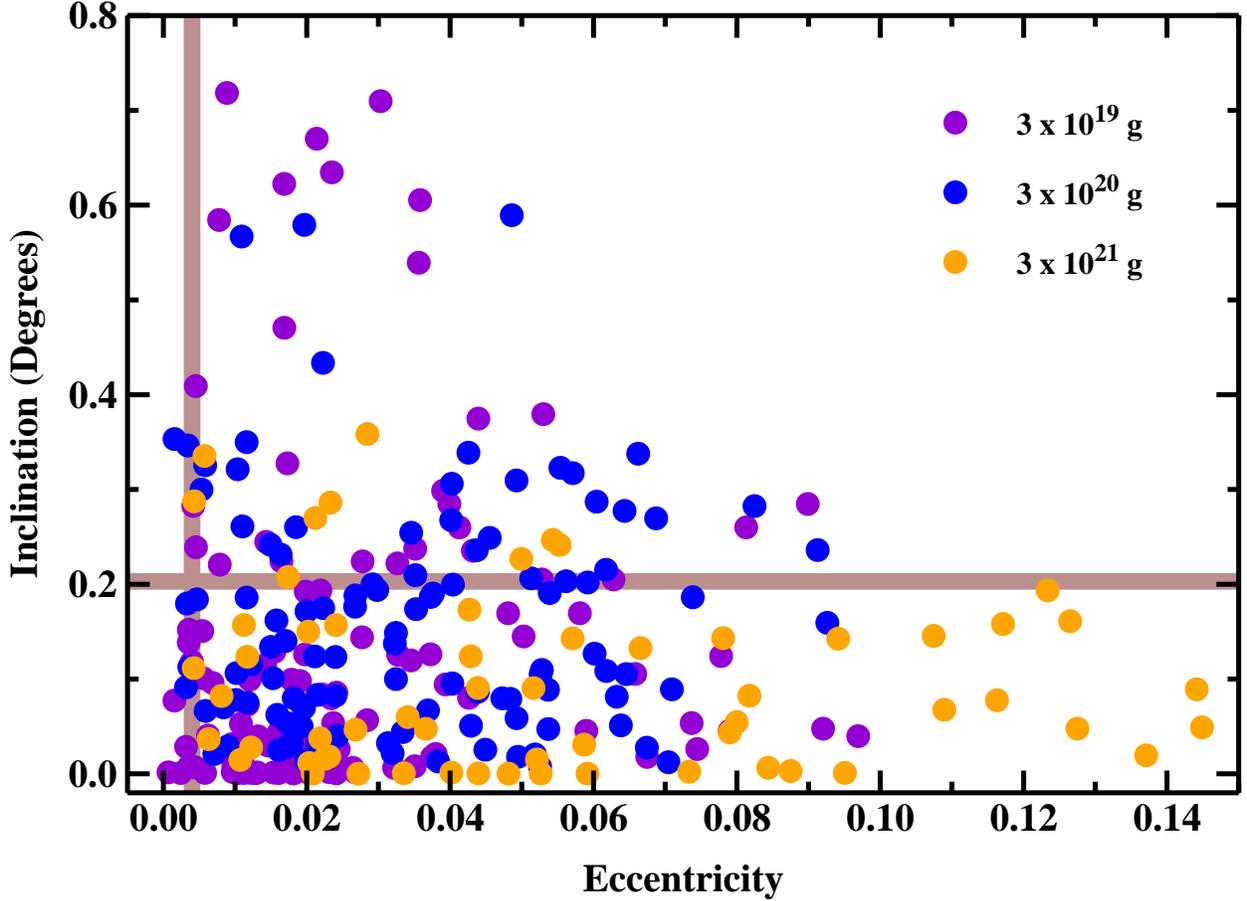}
\vskip 3ex
\caption{
Distribution of orbital eccentricity $e$ and inclination $i$ for 
satellites produced in disks with initial masses of 
$3 \times 10^{19}$~g (violet points), 
$3 \times 10^{20}$~g (blue points), and
$3 \times 10^{21}$~g (orange points). 
Although there is no clear correlation of average orbital 
inclination with disk mass, disks with larger initial masses 
produce satellites with larger average $e$. 
The horizontal and vertical bars indicate the average of the
measured eccentricity and inclination for Nix ($e_N$ and $i_N$)
and Hydra ($e_H$ and $i_H$) from the 2007 Jacobson PLU017 JPL 
satellite ephemeris. For the model satellites, 10\% have 
$e \lesssim e_{H,N}$ and  83\% have $i \lesssim i_{H,N}$.
\label{fig: ei}
}
\end{figure}

To conclude this section, Fig.~\ref{fig: ei} plots pairs of $(e, i)$ for satellites that survive
for $10^7$ yr. For most satellites, the orbital inclination is small: the average inclination is
$<i>$ = 0.1\degree\ $\pm$ 0.2\degree. Despite the apparent excess of high $i$ objects at small $e$ 
for the low and intermediate mass disks, the average $i$ and the dispersion in the average do not 
vary with initial disk mass.  
The average $e$ grows with disk mass, however, with
$<e> = 0.02 \pm 0.02$ for $M_d = 3 \times 10^{19}$~g, 
$<e> = 0.03 \pm 0.02$ for $M_d = 3 \times 10^{20}$~g, and
$<e> = 0.06 \pm 0.06$ for $M_d = 3 \times 10^{21}$~g. 
For small and intermediate initial disk masses, the average eccentricity is very low.
For large disk masses, the $<e>$ is a factor of 2--3 larger. 

Although the final orbital elements of satellites are independent of the promotion mass, 
$e$ depends on the initial masses of the largest planetesimals. When we add several large
planetesimals to the initial distribution of 0.1~km and smaller planetesimals, the
calculations yield fewer satellites with larger masses (Fig.~\ref{fig: rsemi2}). As in
calculations with larger total disk masses, a few large satellites stir up their 
surroundings more than many smaller satellites. These satellites then have larger $e$.

The orbital elements of model satellites agree reasonably well with the observed elements 
of the \pc\ satellites.  Roughly 10\% of the model satellites have $e \lesssim e_H$, where 
$e_H$ is the orbital eccentricity of Hydra.  Nearly all (83\%) satellites have orbital 
inclinations $i \lesssim i_H$, where $i_H$ is the inclination of Hydra.  Satellites with 
integer period ratios are also common. For calculations with $M_d = 3 \times 10^{21}$~g,
60\% of the outer satellites have period ratios reasonably close to the 5:3, 2:1, or 3:1 
commensurabilities with the inner satellite. Disks with smaller initial masses produce more
closely-packed, lower mass satellites which often lie close to the 3:2 and 4:3 commensurabilities.
For these lower mass disks, 50\% to 70\% of satellites have period ratios within a few
per cent of the 4:3, 3:2, 5:3, 2:1, or 3:1 commensurabilities.

These results are encouraging. Model satellites often lie close to the observed orbital 
commensurabilities in the \pc\ satellites. The derived inclinations of model satellites 
also agree very well with observations. Although lower mass disks produce satellites with 
the smallest $e$, derived eccentricities for all calculations are factor of 4--20 larger 
than observed. Thus, the calculations have two successes -- the small inclinations and the
likelihood of close orbital commensurabilities -- and one failure -- the lack of very small 
eccentricities for the satellites.

In addition to satellite formation, pure and hybrid coagulation calculations leave behind
remnant disks of small particles. When satellites are close to their final masses, remnant disk 
particles have large orbital eccentricities and large vertical scale heights. At this epoch,
it is more likely for small disk particles to collide with one another than to collide with 
a 10--15~km satellite. A collisional cascade gradually depletes the remnant disk
\citep[e.g.,][and references therein]{kb2008,kb2010}. For conditions in the \pc\ system,
the remnant disk mass at $\sim 10^4$ yr is roughly $10^{15} - 10^{17}$ g and declines with
time as $t^{-1}$ \citep[see also][]{dom2003,kb2004a}.  Thus, the disk has a fairly substantial 
mass for 1--10 Myr.  At later times, the loss of material from collisional grinding reaches
an equilibrium with material captured from collisions between small Kuiper Belt objects 
and the satellites \citep[e.g.][]{stern2006,poppe2011}. Based on current analyses of the 
capture rate, this equilibrium mass is roughly $10^{13} - 10^{14}$ g.

If satellites have much larger radii, $\sim$ 40--70~km, they accrete small particles faster
than the collisional cascade removes them. The remnant disk mass then declines more rapidly, 
on time scales of roughly $10^5$~yr instead of 1--10~Myr. Direct albedo measurements from
{\it New Horizons} will test this possibility.

To explore the outcomes of formation models in more detail, we now consider numerical
calculations of satellite migration through the circumbinary disk surrounding \pc. Aside 
from testing analytic estimates for essential migration parameters, these calculations 
allow us to estimate the frequency of migration-driven mergers and to examine the evolution 
of $e$ and $i$ for an ensemble of migrating satellites. 

\subsection{Migration of Satellites Through the Circumbinary Disk}
\label{sec: sat-mig}

\subsubsection{Background}

Throughout the growth process outlined in \S3.4, gravitational scattering plays an important 
role in the local structure of the disk. The Hill radius,
\begin{equation}
\rhill \approx 10 \left ( {R \over 1~{\rm km} } \right ) \left ( {a \over {\rm 20~\rp} } \right ) ~{\rm km} ~ ,
\label{eq: rhill}
\end{equation}
defines the region where the gravity of a satellite overcomes the gravity of \pc. If the satellite
can clear its orbit of small objects, it can open a gap in the radial distribution of solids.  For 
convenience, we define the Hill radius necessary for a satellite to open up a gap in a dynamically 
cold disk surrounding the \pc\ binary \citep[e.g.,][]{raf2001,bk2013}:
\begin{equation}
\rgap \approx 19 \left ( { v_r \over {75~\cmsec} } \right )^{2/3}
\left ( { M_d \over 10^{20}~{\rm g} } \right )^{1/3}
\left ( { R \over {\rm 1~km} } \right )^{-1/3}
\left ( { a_c \over 20~\rp\ } \right )^{1/3} ~ {\rm km} ~ .
\label{eq: rgap}
\end{equation}
Satellites with \rhill\ $\approx$ 20~km have physical radii $R \approx$ 2~km (eq. [\ref{eq: rhill}]).
For circumbinary disks with $M_d \approx 10^{20}$~g, satellites with $R \gtrsim$ 2~km open gaps in 
the disk.  All four \pc\ satellites have $R \gtrsim$ 3~km. Thus, each can open a gap in a cold 
circumbinary disk.

For satellites with $R \gtrsim R_{gap}$, there are two possible modes of migration
\citep[e.g.,][]{bk2011b,bk2013}. If the satellite can (i) clear a gap in its corotation zone 
and (ii) migrate across this gap in one synodic period, the satellite undergoes fast migration. 
Satellites with $\rhill > \rgap$ ($\rhill = \rgap$) satisfy the first (second) condition. Although 
the size of the gap grows with $R$; the synodic period decreases with $R$. Thus, there is a maximum 
Hill radius -- defined as \rfast -- which allows a satellite to satisfy the second condition.  
With this definition, satellites with $\rgap \lesssim \rhill \lesssim \rfast$ undergo fast 
migration.  For the \pc\ binary,
\begin{equation}
\rfast \approx 20 
\left ( { M_d \over 10^{20}~{\rm g} } \right )^{1/2} 
\left ( { a \over 20~\rp\ } \right )^{3/2} ~ {\rm km}.
\label{eq: rfast}
\end{equation}
Depending on the disk properties and satellite location, this Hill radius corresponds to objects 
with physical radii of a few kilometers. Coupled with the limits on \rgap, satellites with physical 
radii of 2--5~km undergo fast migration. These objects drift radially inward or outward at a rate 
\begin{equation}
\dot{a}_{fast} \approx \pm 5
\left ( { M_d \over 10^{20}~{\rm g} } \right )
\left ( { a \over 20 \rp\ } \right )^{3/2}
{\rm km~yr^{-1}} ~ .
\label{eq: dadt-fast}
\end{equation}
On time scales comparable to the accretion time of 200~yr, 2--5~km objects drift roughly
1~\rp. Thus, growth and migration are simultaneous.

Massive satellites with $\rhill > \rfast$ migrate through the disk at a rate that
depends on the disk viscosity. In gaseous disks with a large viscosity, this ``gap''
migration can transport gas giants from 5--10~AU to within a few stellar radii of a 
solar-type star in $\sim$ 1~Myr \citep{linpap1986,ward1997,nelson2004}. In a disk of 
particles, the smaller viscosity derived from gravitational scattering leads to a smaller 
gap migration rate \citep{ida2000b,cionco2002,kirsh2009,bk2011b,ormel2012}. For a 
circumbinary disk in \pc, the expected migration rate is \citep{bk2013}
\begin{equation}
\dot{a}_{II} \approx -0.15
\left ( { M_d \over 10^{20}~{\rm g} } \right )
\left ( { R \over {\rm 20~km} } \right )
\left ( { a \over 20 \rp\ } \right )^{1/2}
{\rm km~yr^{-1}} ~ .
\label{eq: dadt-type2}
\end{equation}
Satellites with $R \approx$ 10--20~km migrate 1~\rp\ every $10^4$ yr. This rate is
slower than the growth rate. Thus, large satellites grow faster than they migrate.

\subsubsection{Migration of a Single Satellite}

To explore satellite migration in a disk surrounding the \pc\ binary, we use the 
$N$-body component of the \orch\ code.  In this mode, the calculations directly track 
interactions between satellites and massive ``super-particles'' that represent small 
particles in the disk. For massive objects within a disk around a single central 
object,  our calculations \citep{bk2011a,bk2011b} reproduce published results of 
previous investigators \citep[e.g.,][]{malhotra1993,hahn1999,kirsh2009}. Our 
investigation of migration within Saturn's rings includes extensive tests with gaps 
in the disk, orbital resonances, and the gravitational perturbations of distant massive 
moons outside resonance \citep{bk2013}.  To test the theoretical limits on migration 
rates in a disk surrounding \pc, we first consider calculations of a lone satellite 
with $R \approx \rgap - \rfast$ within a particle disk around a single central object 
with the combined mass of \pc.  We then consider how a binary central object modifies 
the mode and rate of migration.  Because the hybrid calculations often produce many 
small satellites, we conclude this section with simulations of multiple satellites 
orbiting a central binary.

For this suite of simulations, we adopt a disk surface density distribution, 
$\Sigma \propto a^{-1}$, with a fixed mass of $3\times 10^{20}$~g. The disk
extends from $a$ = 20 \rp\ to $a$ = 70 \rp\ around a single object with a
radius of 1 \rp\ or a binary with a separation of 5 \rp. We follow an ensemble 
of super-particles and satellites in an annulus of full width of 5 \rp. Super-particles 
have masses of 1/2000$^{\rm th}$ the mass of the satellite.  These objects interact 
with the satellite and the central mass, but not with each other. Satellites have 
fixed bulk mass density, $\rho = 1$~g~cm$^{-3}$. 


At the start of each simulation, super-particles are dynamically cold. Thus, the disk 
is geometrically thin, with $H_z / \rhill \lesssim$ 1.  Particle trajectories evolve 
solely by interactions with the central (single or binary) object and massive satellites. 
Unlike our simulations of Saturn's A ring \citep{bk2013}, there is no collisional damping 
among super-particles.  These initial conditions are ideal to assess a satellite's ability 
to migrate through the disk and lead to fairly robust upper limits on the migration rate.  
In a more realistic disk, collisional damping, dynamical friction, and viscous stirring 
generally produce increases in the velocity dispersion and vertical scale height of disk 
particles on time scales comparable to the growth time (see \S3.4). Because migration 
rates fall off as the inverse cube of the mean disk particle eccentricity 
\citep{ida2000b,kirsh2009,bk2011b}, we expect that migration time scales in a real disk 
are somewhat longer than our estimates for idealized disks.

These simulations assume a constant mass of small particles in the disk. In reality, 
collisions and interactions with the central binary deplete the small particles which
drive migration of proto-satellites.  Shortly after the \pc\ impact, however, collisions 
are destructive. At this epoch, 1--10~km objects which survive the impact may migrate
through the expanding disk. While collisional damping calms disk material, large
survivors first migrate slowly through the disk (eq. [\ref{eq: dadt-type2}]); later, 
they may migrate more rapidly (eq. [\ref{eq: dadt-fast}]).  Once collisions yield 
larger merged objects, small particles deplete on the collision time, which is similar
to the time scale for fast migration. During this phase, newly-formed 1--10~km objects
may migrate rapidly through the disk. As these objects continue to grow, they gradually
open up gaps and deplete the disk of small particles. Because our simulations do not 
currently provide an accurate picture of gap migration with accretion, we restrict our 
calculations to conditions appropriate for fast migration during the early evolution of 
the disk.

\begin{figure}
\includegraphics[width=6.5in]{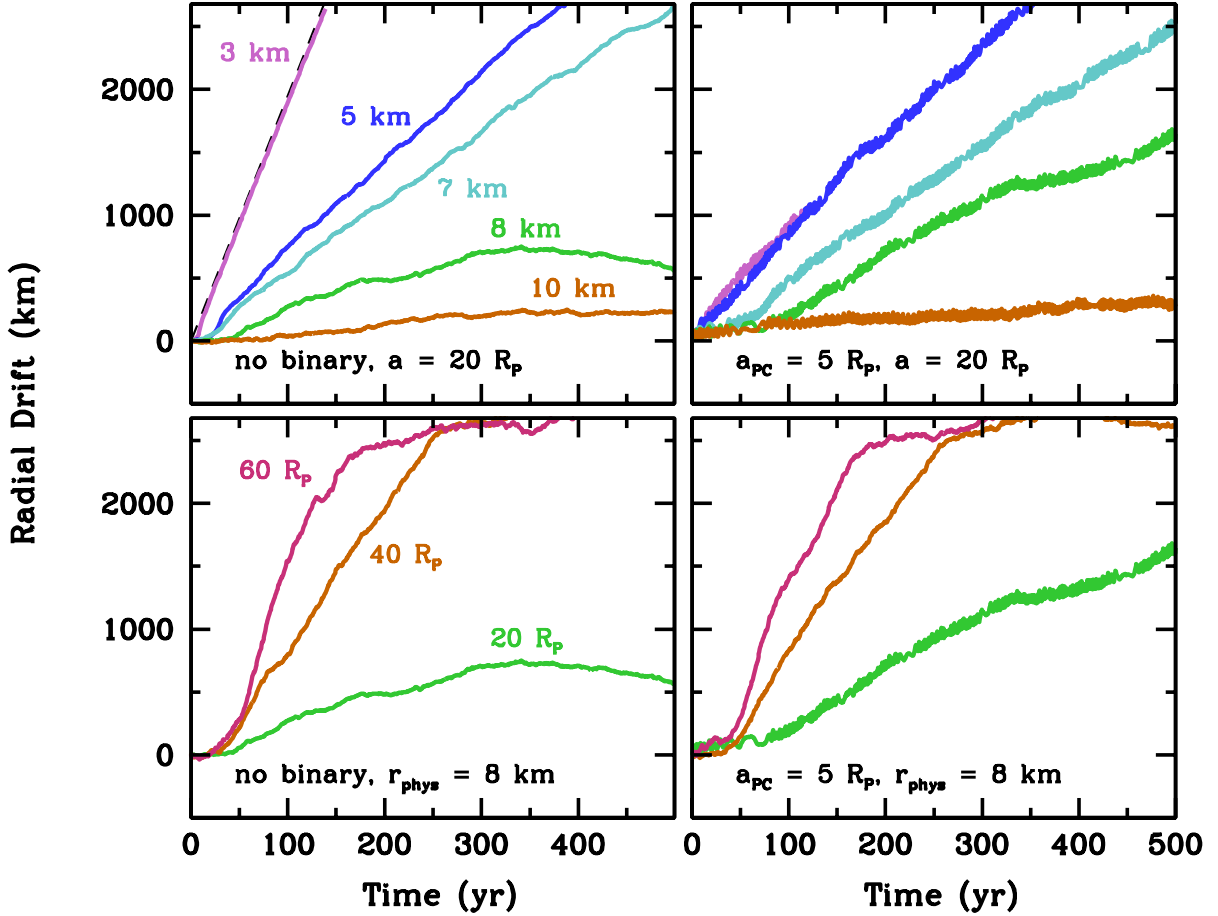}
\vskip 3ex
\caption{
\label{fig: fastmi}
Migration of single satellites through a particle disk.
Each trace shows the result of a simulation with either a
single central point mass (left panels, with mass equal
to the sum of \mp\ and \mc) or a central Pluto-Charon
binary with orbital separation of 5\rp. The legends
give the physical radii of each satellite with $\rho = 1$~g~cm$^{-3}$; 
the legends also show the semimajor axis of the satellites'
initial orbit.  The disk model has a total mass of 
$3\times 10^{20}$~g in a disk with surface density 
$\Sigma \propto a^{-1} $ extending in radius from $20\rp$ to $70\rp$.  
Planets migrate more commonly in the inward direction; however, 
random events can result in outward migration.  
}
\end{figure}

Figure~\ref{fig: fastmi} illustrates results for single satellites around a single central 
mass (left panels) and a central binary system (right panels). At 
$a = 20$ \rp\ (Fig.~\ref{fig: fastmi}; upper left panel), satellites with $R \approx$ 1--4~km 
should open a gap and undergo fast migration (eqs.  [\ref{eq: rgap}--\ref{eq: rfast}]). 
Satellites with $R = 3$~km migrate at rates $\dot{a} \approx$ 20~km~yr$^{-1}$, close
to the rate predicted in eq. (\ref{eq: dadt-fast}). As the satellite radius increases from
3~km to 5--10~km, the migration rate drops considerably. These larger satellites have
too much inertia to maintain the fast radial drift rate. Larger satellites with 
$R \gtrsim$ 10~km migrate at rates close to predicted rates for gap migration 
(eq. [\ref{eq: dadt-type2}]).


Migration rates also depend on the semimajor axis of the satellite 
(Fig.~\ref{fig: fastmi}; lower left panel). When the velocity dispersion of the
disk is fixed, it is harder for smaller satellites with $R$ = 3--4~km to open up a gap 
in the disk at larger $a$ than at smaller $a$ (eq. [\ref{eq: rgap}]).  Thus, small 
satellites migrate much more slowly at large $a$.  However, \rfast\ also increases 
with $a$. Larger satellites migrate more rapidly at larger $a$ than at smaller $a$.  
For a larger satellite with $R =$ 8~km, the migration rate slows considerably from 
$a$ = 60 \rp\ (roughly 20~km~yr$^{-1}$) to $a$ = 40 \rp\ (roughly 10~km~yr$^{-1}$) to 
$a$ = 20 \rp\ ($\lesssim$ 1~km~yr$^{-1}$).  The relative change in the migration 
rate follows theoretical predictions (eqs.~[\ref{eq: dadt-fast}--\ref{eq: dadt-type2}]).

The presence of a central binary changes these results modestly 
(Fig.~\ref{fig: fastmi}; right columns). The most obvious impact of the \pc\ binary
is the increased velocity dispersion of disk particles. Larger velocity dispersion reduces 
the effectiveness of torque exchange between the satellite and disk particles in the 
satellite's corotation zone, making it more difficult for a satellite to open up a gap 
in the disk (eq. [\ref{eq: rgap}]).  Thus, smaller satellites migrate more slowly 
around a binary.  There is some evidence from the simulations that larger satellites
may migrate more rapidly around a binary. In these cases, larger satellites first 
clear their corotation zone and halt their radial drift. As the simulation proceeds,
stirring from the binary feeds some material into the corotation zone, commencing 
a kind of ``attenuated'' migration \citep{bk2011b}. We speculate that this behavior
may allow larger objects to drift more quickly through a circumbinary disk.

Simulations with a variety of disk masses confirm these general trends. Under the 
right conditions in disks with masses larger than $\sim 10^{18}$~g, satellites with 
$R \approx$ 1--20~km can undergo fast, gap, or attenuated migration with rates 
ranging up to 20~km~yr$^{-1}$. In lower mass disks, small satellites cannot clear the
gap required to initiate fast migration. 
Although large satellites can form gaps in disks with $M_d \lesssim 10^{18}$~g, 
the minimum radius required to open a gap is larger than the maximum radius for 
fast migration. Thus, large satellites cannot migrate in the fast mode. Instead,
these objects migrate a factor of ten more slowly in the gap migration mode.


To test whether the disk can circularize the orbits of small satellites, we derive
predicted rates for 4--10~km objects in disks with $M_d = 1 - 10 \times 10^{17}$~g
surrounding a single central object.  For satellites in this size range, damping 
rates are independent of radius. The damping time scales with the disk mass:
\begin{equation}
t_{damp} \approx {e \over \dot{e}} \approx 10^5 \left ( {10^{17}~{\rm g} \over M_d} \right ) ~ {\rm yr} ~.
\label{eq: dedt}
\end{equation}
Even in a low mass disk, damping times are comparable to the formation time for 4--10~km
satellites.

Estimating damping rates for satellites orbiting a central binary is complicated by 
precession, orbital resonances, and other dynamical issues. For the small rates
indicated by simulations with a single central object, accurate estimates require a
substantial investment of cpu time. When the \pc\ binary has a circular orbit, 
dynamical friction between the satellite and disk particles will still circularize the 
satellite's orbit. In an eccentric \pc\ binary, the damping time probably depends
on the time scale for the disk and tides to circularize the binary. Because these 
processes act on similar time scales, deriving the circularization time for a satellite
orbit is very cpu intensive and beyond the scope of this study. 

\subsubsection{Migration of Multiple Satellites}

\begin{figure}
\includegraphics[width=6.5in]{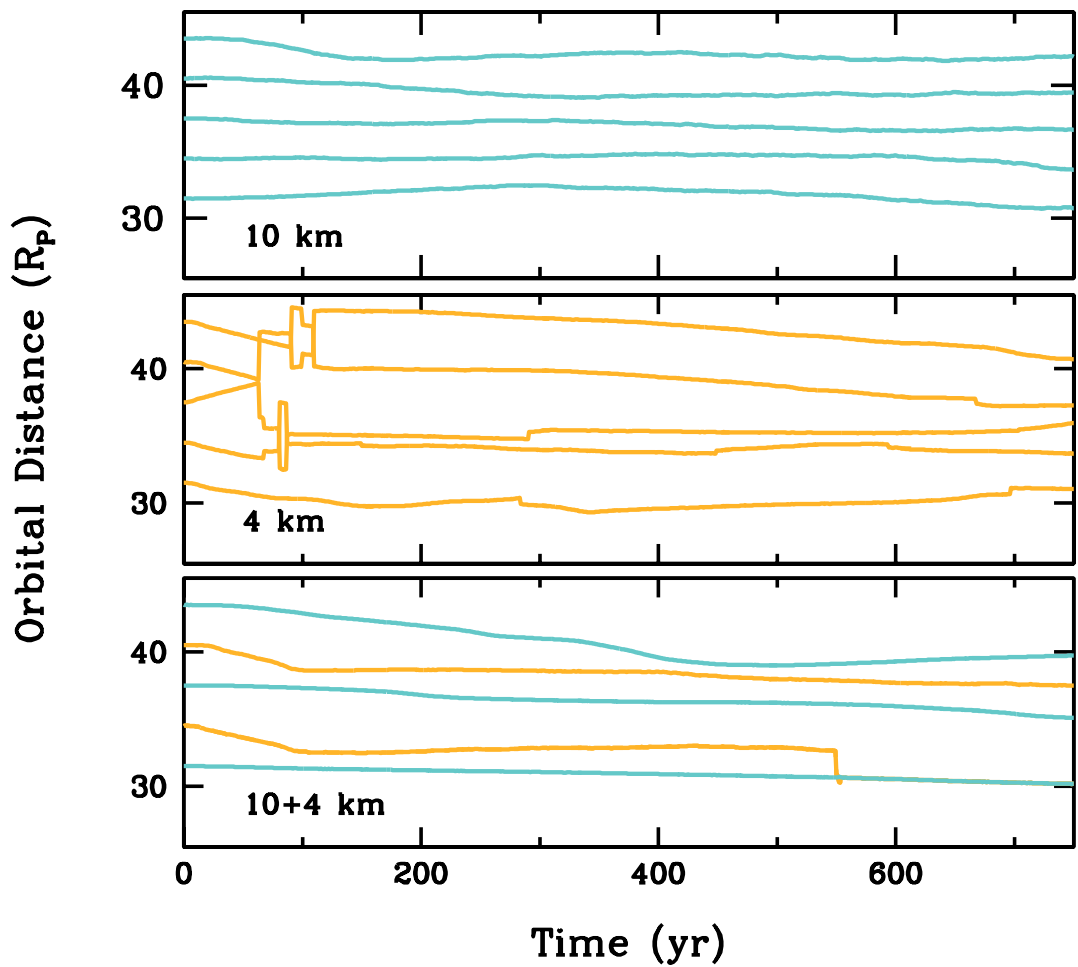}
\vskip 3ex
\caption{
\label{fig: multixx}
Migration of multiple satellites around a tight Pluto-Charon binary.
The binary orbit and disk model are the same as in Figure~\ref{fig: fastmi},
except that each set of five satellites is embedded in a 20\rp\ annulus 
of disk particles. Each panel gives an example of the simulation
output; the upper panel shows equal-mass 10~km satellites, while the
middle panel tracks smaller 4~km bodies. In the former case, some
radial drift occurs but does not lead to significant scattering or merging.
The smaller satellites are more interactive; the middle panel shows an orbit
crossing event. The lower panel illustrates differential migration in which
a smaller body merges with a larger one.
}
\end{figure}

To investigate whether ensembles of proto-satellites can migrate, we examine a second 
suite of simulations (Figure~\ref{fig: multixx}). Here, we maintain the same disk
surface density distribution and focus on satellites with $R$ = 4~km or 10~km orbiting 
within a wide annulus spanning orbital distances of 35\rp\ to 55\rp.  Satellites and
disk material orbit a compact \pc\ binary with orbital separation $a$ = 5~\rp. For
material at 35--55~\rp, stirring by the binary has a smaller impact than at 20~\rp.

The top panel of Figure~\ref{fig: multixx} follows a simulation of five 10~km 
satellites initially evenly spaced in orbital separation. With typical Hill radii of
0.2~\rp, these satellites interact weakly among themselves. All have radii larger
than \rfast; migration rates are slow.  Modest scattering of disk particles yields
some inward or outward motion and occasional stronger gravitational interactions among
the satellites. However, these satellites are fairly stable over 1000~yr and longer 
timescales.

Simulations with ensembles of five 4~km objects produce more complicated outcomes.  With 
much smaller Hill radii, these satellites interact very weakly among themselves. 
At the onset of each simulation, however, all begin to migrate in the fast mode. Some
migrate inward; others migrate outward. When neighboring satellites migrate in the same
direction, migration is limited by the stirred up wake of disk particles left behind by 
the first satellite to pass through that portion of the disk \citep{bk2011b}. Satellites 
do not cross these wakes. Thus, groups of small satellites migrating in the same direction 
drift radially inward/outward a small distance before the migration stalls.

Small satellites migrating in opposite directions lead to more interesting outcomes.
Often, these satellites scatter one another, sometimes exchanging orbits 
(Fig.~\ref{fig: multixx}; middle panel). Depending on the sense of migration and the
state of the disk after scattering, satellites may undergo multiple scattering events 
before settling down in roughly stable configurations within a stirred up disk. 
Sometimes, satellites merge.  Because merged satellites are massive, they migrate more
slowly and end up in a stable configuration. In our suite of simulations, mergers are
less common ($\sim$ 10\%) than scattering events.

Simulations with a mix of 4~km and 10~km satellites also lead to interesting outcomes
(Fig.~\ref{fig: multixx}; lower panel). If a small satellite undergoing fast migration 
lies between two larger satellites, its larger migration rate guarantees that it will 
catch up to one of its slowly moving neighbors. When it does catch up, it may `bounce off' 
the wake of its neighbor and begin migrating in the opposite direction. In roughly half 
of all interactions, the larger satellite either scatters or merges with the smaller 
satellite. If scattering places the small satellite on an orbit with a pericenter 
inside $a_{min}$, the central binary can eject the small satellite out of the system. 


\subsubsection{Summary of Migration Calculations}

For satellites with $R \approx$ 4--10~km in a disk surrounding the \pc\ system,
migration is ubiquitous.  In a cold, geometrically thin disk, satellites in this 
size range can undergo fast mode or slower, gap migration at rates very close to 
those predicted by analytic theory. These rates lead to significant radial motion, 
$\sim$ 1--10 \rp, on timescales comparable to the growth time of 1--10~km satellites, 
$\sim 10^2 - 10^3$ yr. 

Interactions between the disk and the central binary can augment or reduce migration
rates. Although changes to the rates are modest, smaller (larger) satellites generally
migrate more slowly (rapidly). Changes are more significant closer to the binary, 
where the jitter of particle orbits in the disk modifies the number of particles in
the corotation zone of a satellite. Thus, small (large) satellites which form closer 
to the binary are less (more) likely to migrate than satellites which form farther
away from the binary.

Ensembles of migrating satellites produce a variety of interesting dynamical phenomena.
In systems with several small satellites or a mix of large and small satellites,
differential migration enables large-scale scattering events and satellite mergers. 
When scattering places a satellite on a orbit with a pericenter smaller than $a_{min}$,
the central binary ejects the satellite from the system. Mergers among migrating
satellites allow lower mass disks to produce more massive satellites.

The ability of satellites to migrate through an evolving disk depends on the ratio 
of the vertical scale height $H_z$ to the Hill radius $\rhill$.  When collisional 
damping dominates particle stirring, $H_z / \rhill \lesssim$ 1. Satellites can 
open gaps in the disk and migrate at the nominal rates.  In a disk dominated by 
particle stirring, $H_z / \rhill \approx 3 (a / 20 \rp)^{1/2}$.  When 
$H_z / \rhill \gtrsim$ 1, satellites cannot open gaps in the disk. Migration is then 
`attenuated,' with a rate smaller by a factor of roughly $(H_z / \rhill)^3 \approx$ 
30--100 at $a \approx$ 20--60 \rp\ \citep{ida2000b,kirsh2009,bk2011b}. Despite this 
reduction, the attenuated gap migration rate is still substantial, $\sim$ 1 \rp\ every 
10 Myr, similar to the time scale for tides and dynamical interactions to modify the 
semimajor axis of the binary orbit.

The ability of satellites to migrate efficiently also depends on the depletion 
rate of small particles in the disk. Small particles drive the migration of large 
particles.  As particles grow from 0.1~km and smaller objects into 1~km and larger 
objects, the mass of the disk capable of driving migration declines.  When the
largest objects have $R \approx$ 1--5~km, most of the disk mass is in small objects.
Thus, the large objects migrate at the nominal rate. Once the largest objects reach
sizes of 10--30~km, small objects contain less than half of the initial disk mass. 
Migration is then slower than the nominal rate.

Scaling the migration rates derived in the simulations to the hotter disks expected 
from the coagulation and hybrid calculations in \S3.4, satellites with radii similar 
to Kerberos and Styx may migrate significantly as they grow. Although larger satellites 
such as Nix and Hydra migrate more slowly, radial motion of a few Pluto radii seems 
likely after they reach their final sizes.

Although remnant particle disks with low masses, $\sim 10^{17}$~g, cannot drive 
migration on short time scales, they are effective in circularizing the orbits of 
satellites with sizes of 1--10~km.  The time scale to circularize the orbit of a
satellite is a factor of 10--100 times shorter than the time scale for tides to 
expand the orbit of the \pc\ binary. Large final eccentricities, $e \approx$ 
0.02--0.1, are a common feature in hybrid simulations of satellite formation (\S3.4). 
Once satellites reach their final masses on time scales of $10^4 - 10^5$ yr,
remnant disks with masses of $10^{16} - 10^{17}$~g are ubiquitous. Thus, the 
migration simulations demonstrate that satellites embedded in a low mass disk 
will have the small eccentricities observed in the \pc\ system.

This discussion suggests two plausible phases for migration in a circumbinary
disk surrounding \pc. During the early stages of the evolution, rapid migration of 
small satellites is possible when collisional damping dominates particle stirring.
During this phase, growing satellites can migrate at least several Pluto radii.
Later on, when stirring by newly-formed satellites dominates collisional damping,
attenuated migration in a low mass disk slowly modifies satellite orbits as the 
inner binary expands. During this period, satellites may migrate much less than
a Pluto radius, especially if the collisional cascade removes most of the small 
particles remaining in the disk.

\section{DISCUSSION}
\label{sec: disc}

Our analysis in \S3 suggests that the steps outlined in \S2 provide a reasonable path to the formation 
of small satellites close to their current positions within debris from the giant impact.  
This process begins (step 1) with the growth of Pluto-mass planets in a circumstellar disk of 
solid material.  Once Plutos form, giant impacts are common (step three).  During intervals 
between impacts (step two), Plutos capture $\sim 10^{19}$~g of debris. The debris has large $a$ 
and probably does not interact with material from the impact.  Following impact (step 4), material 
ejected from the impact lies at $a \approx$ 20~\rp. In a reasonably massive disk composed of 
0.1--1~km particles, angular momentum transfer from the binary to the ring and among ring 
particles spreads the ring to the current positions of the satellites on fairly short time 
scales.  As the ring spreads, high velocity collisions diminish the sizes of the largest 
particles and remove mass from the ring. Eventually, collisional damping overcomes secular 
perturbations from the binary, enabling the growth of satellites within a broad ring. As 
satellites grow, they migrate through the ring. Although our calculations do not yield 
satellites at the same positions as Styx--Hydra, they yield satellites at similar positions.
Often, stable satellites lie in approximate resonances.

Despite the plausibility of this picture, there are definite uncertainties. During the interval
prior to impact, Pluto and Charon probably capture a modest amount of material. However, it is 
not clear whether any of this material can interact with debris from the impact. After impact, 
there is a broad range of likely values for the binary eccentricity and the properties of the 
ejecta. Within this range, it is not certain that collisions can damp velocities and redistribute 
angular momentum fast enough to overcome secular stirring, spread the ring into a disk, and 
enable the growth of smaller particles. Although it is plausible that a few large particles with 
$R \approx$ 1--10~km escape destruction during ring formation and serve as the seeds for large
satellites (\S3.4), they may be unable to flow out in an expanding ring of smaller objects. 
Even if satellites form within material in an expanded disk of material, they must survive the
expansion of the \pc\ binary. For satellites at $a \lesssim$ 30~\rp, survival seems unlikely
\citep{peale2011}. To survive at larger $a$, satellites must avoid interactions that destabilize
their orbits.

Once a stable disk forms at $a \approx$ 15--70~\rp, our calculations yield testable predictions 
for the masses and orbital configurations of the \pc\ satellites.
The timescale for satellite growth depends on the properties of the binary and the ring of debris.
In any configuration, there is a period where collisional damping must overcome tidal forcing from 
the central binary. Although we have not calculated this evolution in detail, our estimates in 
\S3.3 imply a time scale of at least 1--10~yr. As this evolution proceeds, the ring probably loses 
a significant fraction of its initial mass.  Once collisions calm the disk, the formation time for 
1--10~km satellites is 
\begin{equation}
t_f \approx 10^3 \left ( {3 \times 10^{19} ~ {\rm g} \over M_d } \right ) \left ( { a \over 30 \rp } \right )^{5/2} ~ {\rm yr}.
\label{eq: t-form}
\end{equation}

During oligarchic and chaotic growth, proto-satellites radially migrate through the 
leftover debris.  Although proto-satellites cannot fall into the \pc\ binary, 
satellites can traverse a substantial fraction of the circumbinary disk throughout 
chaotic growth. Migration can induce mergers among lower mass satellites, enabling 
more massive satellites to form in lower mass disks. Ensembles of satellites with 
similar masses migrate in step with one another, enabling satellites to find stable 
equilibrium orbits around \pc.

Migration rates depend on the nature of the central object. Stirring by a central 
binary can heat the particle disk and slow migration of satellites with radii of 
$O(1)$~km. The binary may also increase migration rates of larger satellites. 
Migration is driven by the asymmetry of inward and outward satellite-disk scattering 
events. Compared to a single central mass, a binary preferentially disturbs a larger
fraction of the orbits of the inwardly scattered disk particles and prevents their 
return to the satellite.  We plan additional simulations to explore whether this 
possibility is sufficient to produce the derived changes in migration rates.

Migration rates are also sensitive to the mass of small objects remaining in the 
disk. When satellites reach sizes of 1--10~km, small particles still contain most 
of the mass. If collisional damping can maintain a small vertical scale height,
satellites migrate 1--10~\rp\ through the disk. As these satellites grow, they
migrate more and more slowly due to the larger vertical scale height and smaller
total mass in small particles. Eventually, migration stalls completely.  Throughout 
this process, interactions between the satellites and the remnant disk of small 
particles circularize the orbits of the satellites.

Together with previous results from \citet{youdin2012}, our calculations 
indicate several conclusions about each satellite. 

\begin{itemize}

\item The semimajor axis of Styx is almost as close as possible to the minimum stable 
circumbinary semimajor axis $a_{min}$. In our framework, Styx has a low mass because
(i) the debris ring had less mass close to \pc\, (ii) it was scattered out of 
a more massive region of the ring by Nix or Hydra, or (iii) after formation 
at large $a$, it migrated into an innermost stable orbit.

\item The orbit of Kerberos lies within a small stable region between the orbits of 
Nix and Hydra \citep{youdin2012}. More massive satellites in this region are
unstable. From the coagulation and migration calculations, Kerberos is a smaller 
remnant of the accretion process.

\item The masses of Nix and Hydra are consistent with the coagulation models.
The central cores of either satellite might have survived the giant impact and then 
accreted debris leftover from the impact. Alternatively, the satellites could form 
in a disk of small particles orbiting \pc.

\end{itemize}

Aside from informing the next generation of numerical simulations for satellite
formation, the {\it New Horizons} mission can test several predictions of our 
calculations. 

\begin{itemize}

\item In our simulations, more massive disks generally produce fewer small satellites.
Forming four (or more) satellites orbiting \pc\ implies circumbinary disks with
relatively small masses, $M_d \lesssim 3 \times 10^{20}$ g. Together with numerical 
simulations of the orbital stability of the satellites \citep{youdin2012}, these 
results require satellite albedos, $A \approx$ 0.4--1. This range for $A$ is similar 
to the measured $A \approx$ 0.5 for Pluto and Charon 
\citep[see][and references therein]{marcialis1992,roush1996,brown2002,buie2010a,buie2010b} 
and much larger than the $A$ for many Kuiper belt objects 
\citep[e.g.,][]{stansberry2008,brucker2009}.  Direct measurements of $A$ by 
{\it New Horizons} will test these constraints.

\item All of the hybrid calculations leave several very small satellites orbiting 
within an extended disk of debris beyond the orbit of Hydra.  The radii of typical 
leftovers -- $R \lesssim$ 1--3~km -- imply optical magnitudes of 28 or fainter. 
Testing this prediction is a challenge for {\it Hubble Space Telescope} 
\citep{showalter2011, showalter2012}, but satellites this size or smaller are 
easily visible during the flyby of {\it New Horizons}. For objects with 
$R \approx$ 1--3~km, the most likely range of semimajor axes is $a \approx$ 
70--90 \rp. Smaller satellites might have $a \approx$ 100--200 \rp. Both sets 
of small satellites should have small inclinations relative to the \pc\ orbital 
plane, similar to the observed $i$ for the known satellites. If smaller particles 
in a leftover debris disk (see next item) have sufficient mass, these outer 
satellites should also have circular orbits.  Smaller masses in leftovers imply 
larger $e$ for small outer satellites. We plan numerical calculations to test the 
longevity of these satellites in the tidal field of the Sun--Neptune system.

\item Estimates for the spreading time in \S3.3 suggest scattering will produce 
ensembles of smaller particles, $R \lesssim$ 0.01--10~m, in an extended disk well 
beyond the orbits of the known satellites.  For a mass of $\sim 10^{15}$~g in 
leftover small particles with typical radius R, the predicted optical depth of 
the disk is $\tau_l \sim 10^{-10} R^{-1}$ at semimajor axes $a \approx$ 70--200~\rp. 
Larger (smaller) disk masses imply proportionately larger (smaller) optical depths.  
Although this mass is significant, the predicted optical depth is much smaller than 
the $\tau \lesssim 10^{-7}$ estimated for clouds of 1--100~$\mu$m particles produced 
by impacts of small Kuiper belt objects on the known satellites \citep{poppe2011}. 
{\it New Horizons} measurements of the optical depths of small particles surrounding 
the known satellites and at larger distances should place interesting constraints on 
the long-term evolution of the satellite system and its interactions with the Kuiper belt.

\item Measuring the amount of material at $a \gtrsim$ 100~\rp\ tests formalisms for the
capture frequency (\S3.2).  In our estimates, most of the mass lies at large semimajor 
axes, $a \approx$ 1000~\rp.  Failure to identify any material at 
$a \approx$ 100--1000~\rp\ suggests that capture is less frequent than our estimates.
In principle, the amount of material at these distances provides a measure of the 
capture efficiency.

\item Composition data from {\it New Horizons} should also test the giant impact 
scenario.  If the relative contamination from captured material (\S3.2) or impacting 
Kuiper belt objects \citep[e.g.][]{stern2006,poppe2011} is small, satellites should 
have roughly the same composition as Charon.  The broad variety of physical 
characteristics among Kuiper belt objects \citep[see][]{barucci2008} suggests that 
satellites with significant contributions of material from the Kuiper belt should 
have very different compositions from satellites grown from the debris of a giant 
impact \citep[see also][]{stern2009}.  Thus, direct measurement of the compositions 
of individual satellites and gradients in the composition among any smaller 
satellites and debris should constrain scenarios for satellite formation.

\item Observed shapes of satellites might also test formation models. In general, 
large fragments from the impact should be more irregular than satellites grown
from a circumbinary disk/ring of much smaller particles 
\citep[see][and references therein]{tanga2009a,tanga2009b}.

\end{itemize}

\begin{figure}
\includegraphics[width=6.5in]{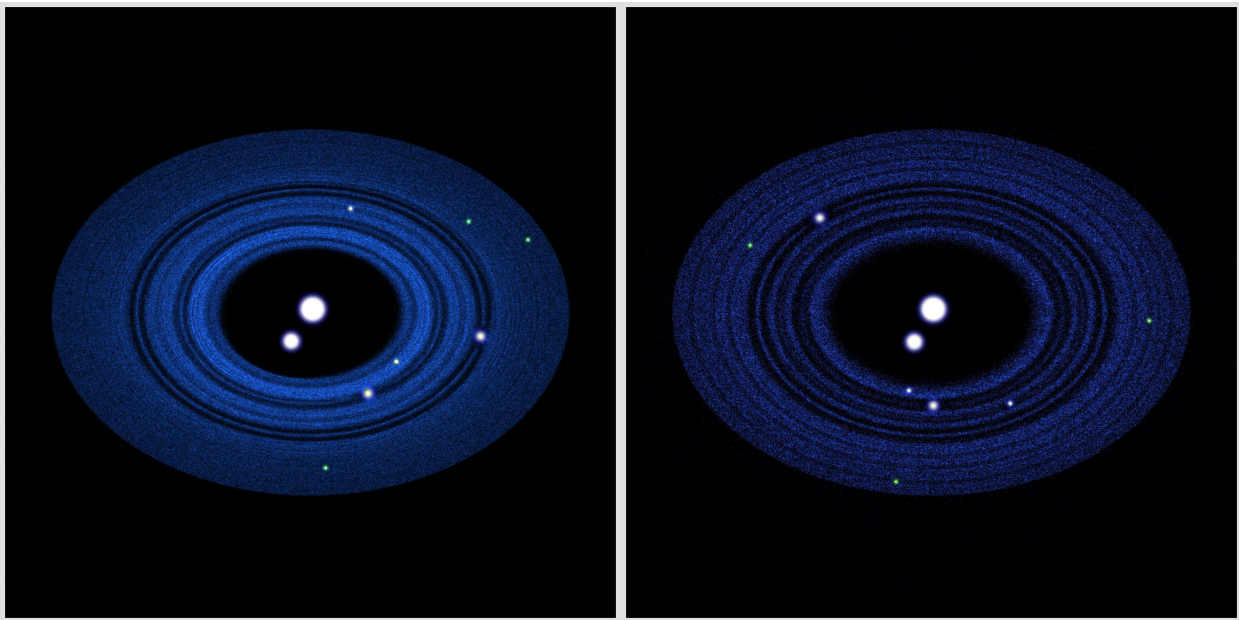}
\vskip 3ex
\caption{
\label{fig: config}
Predicted configuration for the \pc\ system.  The \pc\ binary (represented 
by the two largest white disks), the four small satellites -- Styx, Nix, 
Kerberos, and Hydra (represented by the four small white disks), and 
three smaller satellites (represented by the green disks) lie within 
an extended ensemble of solid particles shown as small blue dots.  
Left panel: System configuration after $10^2$ yr of a computer simulation 
with two million massless tracer particles surrounding the known and 
predicted moons. 
Right panel: System configuration after $10^4$ yr of a computer simulation 
with 0.5 million massless tracer particles.
On short times scales, satellites remove tracer particles along their orbits.
On long time scales, satellites remove nearly all tracers in the inner disk;
low mass satellites in the outer disk begin to clear their orbits.
On time scales of $\gtrsim 10^6$ yr, satellites will clear the inner disk
of small particles \citep[see also][]{youdin2012}; collisional erosion and
scattering will probably remove most of the particles in the outer disk.
}
\end{figure}

To develop predicted configurations for the \pc\ system during a {\it New Horizons}
encounter, we ran $n$-body simulations of \pc, the known small satellites, and 
several predicted satellites embedded within a disk of $0.5 - 2 \times 10^6$ massless 
tracer particles.  On $10^2$ yr time scales (roughly 5000 orbits of the inner binary), 
the larger satellites clear many tracers from their orbits (Fig. \ref{fig: config},
left panel).  
On $10^4$ yr time scales, the four inner satellites clear most of the tracers from
the inner disk (Fig. \ref{fig: config}, right panel).  Although the three outer 
satellites can clear tracers along their orbits, most tracers remain between satellite 
orbits in the outer disk.

On much longer (1--10~Myr) time scales, the inner satellites clear their orbits 
completely \citep[see also][]{youdin2012}. In the outer disk, a few small satellites
with radii of a few km cannot clear more than a few hundred km around their orbits. 
Thus, many tracers remain.  The predicted optical depth and extent of the disk at
late times depends on (i) the number of small satellites beyond Hydra, (ii) the 
outcomes of collisions between small disk particles, and (iii) the amount of mass 
in small particles produced from recent impacts on the satellites 
\citep[e.g.,][]{stern2009,poppe2011} and capture from the Kuiper Belt.  
If (i) there are few small satellites, (ii) collisions damp and do not destroy
small disk particles, and (iii) captures and impacts are common, we expect 
{\it New Horizons} to detect a modest disk of small particles with rings and gaps 
similar to those shown in the right panel of Fig~\ref{fig: config}). If, however, 
(i) there are many small satellites beyond Hydra's orbit, (ii) collisions destroy 
small disk particles, and (iii) captures and impacts are relatively rare, the 
{\it New Horizons} encounter may reveal little or no disk material. Once 
{\it New Horizons} has mapped the \pc\ system completely, improved numerical 
simulations will enable a more complete examination of the formation and history 
of the satellite system.

\section{SUMMARY}
\label{sec: summary}

We describe a theoretical framework for the origin of Pluto's small satellites in the 
context of a giant impact that produces the \pc\ binary \citep[e.g.,][]{canup2011}.
Our main results follow.

\begin{itemize}

\item Throughout the early history of the solar system, giant impacts capable of
producing the \pc\ binary have a frequency of 1 per 100--300 Myr.  These impacts 
leave behind enough debris to enable satellite formation \citep{canup2011}. Once
the debris lies within a ring, angular momentum transport among ring particles 
may enable the ring to spread into a disk.

\item In a ring or a disk of small particles, $R \lesssim$ 0.1~km, surrounding \pc, 
satellites with $R \approx$ 10--80~km form rapidly on a time scale of $10^3 - 10^4$~yr. 
The formation time is similar to the migration time scale.

\item After the impact, objects with $R \approx$ 1--10~km might survive collisional 
disruption during the ejection process. Calculations of disks with small particles
and a few seeds produce fewer satellites with larger masses than calculations with
small particles alone.

\item Radial migration encourages mergers of small satellites within an evolving
disk surrounding \pc. 

\item Scattering events among differentially migrating satellites sometimes place
objects on orbits with pericenters close to the central binary. Unless interactions
with disk particles can damp the orbit, the \pc\ binary ejects these satellites from
the system.

\item For identical masses of the central planet(s), the surrounding disk, and a small
satellite embedded in the disk, migration rates around a binary planet differ slightly 
from those around a single planet. For \pc, the drift rate of a satellite the size of
Styx may be as high as 5~km~yr$^{-1}$.

\item The current number of small \pc\ satellites strongly favors low mass disks
\citep[see also][]{youdin2012}. In our calculations, disks with masses 
$M_d \approx 3 - 10 \times 10^{19}$~g have the best chance at producing 3--5 
satellites with orbital semimajor axes $ a \approx$ 40--60~\rp. Matching observations
requires Pluto's satellites to have albedos $A \approx$ 0.4--1. {\it New Horizons}
data will test this prediction.

\item Model satellites produced in formation calculations have orbital $i$ close to 
those observed, but the typical $e$ is generally too large. Orbital migration can
reduce $e$ to acceptable levels.  

\item The calculations predict a few very small, $R \lesssim$ 1--3~km, satellites and
an extended disk of even smaller particles, $R \approx$ 1--100~cm, beyond the current
orbit of Hydra. Detecting these satellites and the disk from the ground is very 
challenging. If they are present, {\it New Horizons} should detect them easily.

\end{itemize}

Our results demonstrate that numerical calculations can produce simulated satellite
systems with properties reasonably close to those observed. Prior to the {\it New Horizons} 
flyby of \pc, we expect to refine the predictions considerably. Once {\it New Horizons} 
probes the masses, orbital architecture, and composition of the \pc\ binary, a rich 
interplay between the data and the numerical simulations will enable a much more robust 
theory for the formation of satellites (planets) in binary planetary (stellar) systems.

\vskip 6ex

We acknowledge generous allotments of computer time on the NASA `discover' cluster and
the SI `hydra' cluster.  Advice and comments from M. Geller and A. Youdin greatly improved 
our presentation.  We thank two anonymous referees for thoughtful comments that honed our 
analysis. Portions of this project were supported by the {\it NASA } {\it Astrophysics Theory} 
and {\it Origins of Solar Systems} programs through grant NNX10AF35G and the {\it NASA} 
{\it Outer Planets Program} through grant NNX11AM37G.

\bibliography{ms.bbl}

\clearpage

%
\clearpage

\end{document}